%% file: WolfgangMemorial.tex
\documentclass{ws-rv9x6}
\usepackage{ws-rv-van}             
\makeindex

\usepackage{xcolor}

\include{macros}

\begin{document}

\chapter{Wolfgang Kummer and the Vienna School of Dilaton (Super-)Gravity}

\author[L.~Bergamin and R.~Meyer]{Luzi Bergamin\footnote{Email: \texttt{bergamin@tph.tuwien.ac.at}} and Ren\'e Meyer\footnote{Email: \texttt{meyer@mppmu.mpg.de}}}

\address{\footnotemark[1] ESA Advanced Concepts Team, ESTEC, DG-PI,\\
Keplerlaan 1, 2201 AZ Noordwijk, The Netherlands.}

\address{\footnotemark[2] Max-Planck-Institut f\"ur Physik (Werner-Heisenberg-Institut),\\
F\"ohringer Ring 6, 80805 M\"unchen, Germany.}

\begin{abstract}
Wolfgang Kummer was well known for his passion for axial gauges and for the formulation of gravity in terms of Cartan variables. The combination of the two applied to two-dimensional dilaton gravity is the basis of the ``Vienna School'', which provided numerous significant results over the last seventeen years. In this review we trace the history of this success with particular emphasis on dilaton supergravity. We also present some previously unpublished results on the structure of non-local vertices in quantum dilaton supergravity with non-minimally coupled matter.
\end{abstract}

\body

\input{part1.tex}

\input{part2.tex}

\input{part3.tex}

\begin{appendix}[Notations and conventions]

\input{appendix.tex}

\end{appendix}

\bibliographystyle{ws-rv-van}

\bibliography{WolfgangMemorial}

\end{document}

%% file: macros.tex
\usepackage{bbm}

\newcommand{\beqs}{\begin{equation*}}
\def\beq{\begin{equation}}

\newcommand{\eeqs}{\end{equation*}}
\def\eeq{\end{equation}}

\newcommand{\beqas}{\begin{eqnarray*}}
\newcommand{\beqa}{\begin{eqnarray}}

\newcommand{\eeqas}{\end{eqnarray*}}
\newcommand{\eeqa}{\end{eqnarray}}

\newcommand{\eq}[2]{\begin{equation} #1 \label{#2} \end{equation}}

\newcommand{\eps}{\varepsilon}

\newcommand{\ga}{\gamma}
\newcommand{\de}{\delta}
\newcommand{\om}{\omega}

\newcommand{\la}{\lambda}

\newcommand{\De}{\Delta}
\newcommand{\Om}{\Omega}

\newcommand{\blist}{\begin{itemize}}

\newcommand{\elist}{\end{itemize}}

\providecommand{\href}[2]{#2}

\DeclareFontFamily{OT1}{rsfs}{}
\DeclareFontShape{OT1}{rsfs}{m}{n}{ <-7> rsfs5 <7-10> rsfs7 <10->rsfs10}{} 
\DeclareMathAlphabet{\mycal}{OT1}{rsfs}{m}{n}

\newcommand{\lrpd}[1]{\overleftrightarrow{\partial_{#1}}}

\def\Det{{\rm Det}}

\def\cV{{\cal V}}
\def\cL{{\cal L}}

\def\cH{{\cal H}}

\def\cM{{\cal M}}

\def\extd{{\rm d}}

\newcommand{\pd}[2]{\frac{\partial {#1}}{\partial {#2}}}

\def\TrL2{{\rm Tr}_{L^2}}

\def\atbdry{\Big|_{\partial \cM}}
\def\atbdry0{\Big|_{\partial \cM_0}}
\def\atbdry1{\Big|_{\partial \cM_1}}

\newcommand{\diff}{\extd}
\newcommand{\gthree}{{\gamma^3}}
\newcommand{\Stext}[1]{\mathcal S_{\mbox{\tiny #1}}}

\def\ve{\varepsilon}

\newcommand{\matp}{\mathfrak{p}}
\newcommand{\matq}{\mathfrak{q}}

%% file: part1.tex
\section{Historical Introduction}

\subsection{Early Attempts to Non-Einsteinian Gravity in 2D}\label{sect:early}

The earliest works by Wolfgang Kummer connected to gravity in two dimensions\cite{Kummer:1991ss,Kummer:1991bt,Kummer:1991bg,Kummer:1991rt} date back to the year 1991, where he realized together with Dominik Schwarz that the Katanaev-Volovich model\cite{Katanaev:1986wk,Katanaev:1990qm,Katanaev:1989mk},\footnote{Here $R_{abcd}$ are the components of the curvature two-form written with tangent space indices only, $T_{abc}$ are the components of the torsion form, and $\mu,\ga,\la$ are constants. In the second line, we expressed everything in terms of the Ricci scalar $R = R^{\mu\nu}{}_{\mu\nu}$ and the Hodge dual of the torsion form, $T^a_{\mu\nu} = \epsilon_{\mu\nu}\tau^a$.}
\beqa
S &=& \int\extd^2x \sqrt{-g} \left[\frac{\mu^2}{2} R_{abcd}^2 + \frac{\ga^2}{2} T_{abc}^2 + \la\right]\,,\label{eq:KatanaevVolovich}\\
&=& \int\extd^2x \sqrt{-g} \left[\frac{\mu^2}{2} R^2 -\ga^2 \tau_a\tau^a + \la \right]\,,\label{eq:KV2}
\eeqa
easily can be solved in an axial gauge using light-cone variables\cite{Kummer:1991ss,Kummer:1991bg} and applying the first order formulation of gravity theories, since the theory exhibits a (nonlinear) Yang-Mills like gauge structure. In that work, the global structure of the solutions, in particular the classification according to their singularities, was discussed as well. Certainly, this insight was based on Wolfgang Kummer's  long-standing experience with noncovariant gauges in non-Abelian gauge theories\cite{Kummer:1974ze}. Furthermore they observed the existence of two branches of classical solutions: A constant curvature branch with vanishing torsion, yielding de Sitter space in the case of \eqref{eq:KatanaevVolovich}, and a nontrivial branch of solutions with both curvature and torsion, which is however labeled by a conserved quantity relating the torsion scalar and the Ricci scalar in a gauge invariant manner. As we will see, this conserved quantity exists in a much larger class of two-dimensional gravity theories and has the interpretation of a quasi-local mass.
 
The noncovariant gauge was then put to good use\cite{Kummer:1991ss,Kummer:1991rt} to show the renormalizability of the Katanaev-Volovich model \eqref{eq:KatanaevVolovich} in a fixed background quantization around flat space, albeit its nonpolynomial interactions. Though these interactions lead to an infinite set of ultraviolet divergent one-loop graphs, only a single quantity is renormalized, such that the renormalized Greens functions reduce to the tree-level graphs. Infrared divergences were shown to be treatable by introducing a small mass regulator, and it was realized that, although no physical S-matrix exists in the model because of the lack of propagating on-shell degrees of freedom, correlators of physical observables such as the Ricci curvature scalar can still yield interesting information.

Based on the idea that the integrability of \eqref{eq:KatanaevVolovich} should be due to an additional dynamical symmetry, it was then realized by Wolfgang Kummer and collaborators\cite{Grosse:1992eb,Kummer:1992af,Grosse:1992vc,Kummer:1992ef} that the Katanaev-Volovich model can be reformulated as a $\mathit{sl}(2,\mathbbm{R})$ gauge theory. They went on to analyze the constraint structure of the theory, and found that the secondary first class constraints form a deformed $\mathit{iso}(2,1)$ algebra, i.e.\ a deformation of the Poincar\'e algebra of $(2+1)$-dimensional Minkowski space. The deformation parameter turned out to be the constant $\gamma$, and sending\footnote{Though the solutions of the classical equations of motion are in some sense singular in this limit\cite{Grosse:1992vc}, it is more interesting than the de Sitter solution, as it has vanishing torsion but nonvanishing curvature ($R^2$ gravity without torsion\cite{Kawai:1993kp}), as can be seen from the first order formulation \eqref{eq:KatanaevVolovichFirstOrder}. For this reason it was named ``Einstein branch''. 
Different limits of the parameters $\mu,\ga,\la$ in \eqref{eq:KatanaevVolovichFirstOrder} yield further 2d gravity models:\cite{Strobl:1993yn} Taking $(\mu,\ga)\rightarrow\infty$ and integrating out $X_a$ afterwards yields 2D BH Gravity\cite{Cangemi:1992bj,Verlinde:1991rf}. The Jackiw-Teitelboim model\cite{Teitelboim:1983ux,jackiw:1984} is obtained in the same limit when rescaling the cosmological constant such that $\la/\ga = \Delta^2$ is kept fixed.} $\ga\rightarrow\infty$ restored an undeformed $\mathit{iso}(2,1)$ symmetry. They found that the above-mentioned conserved quantity\cite{Kummer:1991bg} was one generator of the center of the deformed symmetry algebra, while a second generator vanished on the constraint surface. Wolfgang Kummer investigated the Katanaev-Volovich model minimally coupled to real scalars or fermions\cite{Kummer:1992ef} as well, and found the same dynamical symmetry, albeit the system in general is no longer integrable. Integrability was found to be preserved for chiral solutions, i.e.\ if the fermion has definite handedness, whereby the conserved quantity is just conserved in time but no longer in space, as its spatial divergence is connected to the spatial parts of the chiral current, $\chi_L^\dagger \lrpd \chi_L$. These currents provide a central extension of the algebra of constraints. This result already hinted towards a generalized conservation law including the matter contributions. In Ref.\ \refcite{Kummer:1992ef} it was also mentioned that this dynamical symmetry and the classical integrability might hint towards nonperturbative quantum integrability of the Katanaev-Volovich model, a statement which is in fact correct for a large class of generalized two-dimensional dilaton gravities. 

After so many interesting facets of the Katanaev-Volovich model being found, it might seem that it would already have revealed all of its mysteries. This was, however, not the case. Ikeda and Izawa\cite{Ikeda:1992qz,Ikeda:1993nk} wrote \eqref{eq:KatanaevVolovich} in first order form,\footnote{Here, the dilaton $\phi$ and the Lagrange multipliers for torsion $X^a$ are auxiliary fields, introduced to linearize the square terms in \eqref{eq:KatanaevVolovich}. The notation has been adapted to the one used in \ref{sect:2DG1stOrder}, $T^a = D e^a = \extd e^a + \eps^a{}_b\om \wedge e^b$ is the torsion two-form with spin connection $\omega_{ab} = \epsilon_{ab} \omega$, and according to our conventions\cite{Grumiller:2002nm} the Ricci scalar is $R = 2 \ast \extd \om$.}
\eq{S=\int\limits_{\cM_2}\left[\phi \extd \om  + X_a (D e^a)\right] + \int\limits_{\cM_2} \extd^2 x \sqrt{-g}\left[\la - \frac{\phi^2}{8 \mu^2}  + \frac{X_aX^a}{4\ga^2}\right]\,,}{eq:KatanaevVolovichFirstOrder}
a fact which was then used by Wolfgang Kummer and Peter Widerin\cite{Kummer:1994ur} to further analyze the symmetry structure of the theory:  They found a field-dependent off-shell global symmetry of \eqref{eq:KatanaevVolovich} whose conserved charge is the conserved quantity mentioned above. Furthermore, they showed, driven by the desire to reinterpret the algebra of constraints in a framework not involving the Hamiltonian phase space, that the deformed $\mathit{iso}(2,1)$ symmetry could be re-formulated as a current algebra. The currents were readily identified with the components of the energy-momentum tensor in lightcone gauge, which were shown to fulfill (on compactified space) a Virasoro algebra, similar to the situation in string theory. As will become clear to the reader later, this first order reformulation of a much wider class of gravity theories, together with the choice of lightcone gauge, lies at the heart of the results outlined in this article.

Also the case with minimally coupled bosonic\cite{Katanaev:1994qf} and fermionic\cite{Kummer:1993uk} matter offered some new surprises: A different lightcone gauge, defined in terms of 'lapse' and 'shift' variables, together with the first order formulation \eqref{eq:KatanaevVolovichFirstOrder} of the gravity sector, allowed a counting of free functions in the general solution of the constraints and the equations of motion. It was found that the inclusion of scalars and fermions does spoil integrability of \eqref{eq:KatanaevVolovich}, but only in a mild way, since only one non-trivial first order PDE needs to be solved. 

The combination of Eddington-Finkelstein gauge and first order formulation \eqref{eq:KatanaevVolovichFirstOrder} also played an important role in the nonperturbative quantum treatment of the Katanaev-Volovich model\cite{Haider:1994cw}: In that work Wolfgang Kummer and Florian Haider found the correct path integral measure that allowed to integrate out both the Zweibeine as well as the spin connection. This measure differs from the one used in the perturbative approach\cite{Kummer:1991rt} by a factor of $(-g)^{-3/4}$, arising from the Gaussian path integration over the fields $\phi,X^a$ in lightcone gauge. As different path integral measures correspond in a renormalizable theory to different renormalization schemes with different counterterms, the nonperturbatively useful measure differed from the perturbative one only by a shift in the counterterm present in the quantum effective action\cite{Kummer:1991rt}. Besides reading off the correct measure just from the symplectic structure which can be found directly from the first order form \eqref{eq:KatanaevVolovichFirstOrder}, they also derived the same result from a canonical BRST analysis. We will see in section~\ref{sect:exactpathintegral} that this method reveals best the underlying symmetry structure of the quantum theory, being in essence a nonlinear Yang-Mills theory. Imposing again lightcone gauge the quantum effective action may be calculated exactly and they found that all local quantum effects disappear right away, i.e.\ the counterterms vanish. They also discuss that the same result can be obtained by careful renormalization in the fixed-background quantization of Ref.~\refcite{Kummer:1991rt}, as expected for a gauge-invariant quantum effective action. The Katanaev-Volovich model thus shows no local quantum effects, and all Green functions are determined by their tree-level contributions. 
Nonetheless, the theory is neither classically nor quantum mechanically
trivial since there exists a global degree of freedom\cite{Schaller:1994np}, namely the
quasilocal mass (conserved quantity). Its role becomes even more important
when considering spacetimes with boundary (see Sect.~\ref{sect:boundary}).

\subsection{First Order Formulation of Generalized 2D Dilaton Gravity}\label{sect:2DG1stOrder}

Ref.~\refcite{Haider:1994cw} was the last in a series of papers\cite{Kummer:1991ss,Kummer:1991bt,Kummer:1991bg,Kummer:1991rt,Grosse:1992eb,Kummer:1992af,Grosse:1992vc,Kummer:1992ef,Kummer:1994ur,Kummer:1993uk,Haider:1994cw} devoted to ``Non-Einsteinian Gravity in $d=2$'', i.e.\ the Katanaev-Volovich model \eqref{eq:KatanaevVolovich}. In the subsequent works Wolfgang Kummer and his collaborators\cite{Kummer:1995fk,Kummer:1995qv,Kummer:1995qh,Katanaev:1996bh,Katanaev:1997ni,Kummer:1997si} realized that a much larger class of gravity theories in two dimensions can be treated along similar lines. This step was possible thanks to a generalization of the first order action \eqref{eq:KatanaevVolovichFirstOrder} to
\begin{equation}\label{eq:SFOG}
 \Stext{FOG} = \int_{\mathcal{M}} \bigl( \phi \diff \omega + X^a D e_a + \epsilon \mathcal V(\phi,Y)\bigr)\ ,
\end{equation}
with $Y = X^a X_a/2$, which first appeared in Ref.~\refcite{Strobl:1994yk}. These models, called  First Order Gravities (FOGs), in general do not permit a formulation exclusively in terms of the geometric quantities (curvature and torsion), since the possibility to eliminate the fields $X$ and $X^a$ (which actually represents a Legendre transformation\cite{Strobl:1999wv}) is not guaranteed. However, already in Ref.~\refcite{Katanaev:1996bh} it was then realized that it is preferable to eliminate the torsion dependent part of the spin-connection instead of the dilaton. Remarkably this procedure only involves linear and algebraic equations for any potential $\cV$, which thus may be reinserted into the action without restrictions.\cite{Ertl:2000si} In general, this yields higher derivative theories of gravity. Still, if one restricts to models of type
\begin{equation}\label{eq:GDTpotential}
 \mathcal V(\phi,Y) = V(\phi) + Y U(\phi)
\end{equation}
the ensuing second order action are Generalized Dilaton Theories (GDTs)
\eq{\Stext{GDT} = \frac12\int\extd^2x\sqrt{-g}\left[\phi R - U(\phi)(\nabla \phi)^2 + 2 V(\phi) \right]\,.}{eq:SGDT}
The classical equivalence also holds on the quantum level\cite{Kummer:1997hy}, as long as possible additional matter fields do not couple to torsion, which is the case for scalars and fermions in two dimensions, but for example not for four-dimensional spherically reduced fermions\cite{Balasin:2004gf}.

Analogous to the Katanaev-Volovich model, all matterless FOGs with potential \eqref{eq:GDTpotential} are classically integrable. Their equations of motion\footnote{The $\pm$-indices are lightcone indices in tangent space, cf.~App.~\ref{sec:notation}.}
\begin{gather}
0 = \extd \phi + X^- e^+ - X^+ e^-\,,\label{eq:EOMepm} \qquad
0 = (\extd \pm \om) X^\pm \mp \cV e^\pm + W^\pm\,,\\
0 = \extd\om + \epsilon \pd{\cV}{\phi} + W \,,\qquad 
0 = (\partial \pm \om)e^\pm + \epsilon \pd{\cV}{X^\mp}\,,
\end{gather}
%
in the matter free case, $W=\frac{\de L^{(m)}}{\de \phi}=0$, $W^\pm=\frac{\de L^{(m)}}{\de e^\mp}=0$, can be solved just by form manipulations. In a patch with $X^+\neq0$, the solution is
\beqa\label{eq:FOGlineelem}
 \extd s^2 &=& 2e^+\otimes e^- = 2\extd f\extd \tilde \phi + \xi(\tilde \phi)\extd f^2\,,\\\label{eq:Qw}
 Q(\phi) &=& \int\limits^\phi U(y)\extd y\,,\quad w(\phi) = \int\limits^\phi e^{Q(y)}V(y)\extd y\,,\\\label{eq:Killingnorm}
 \xi(\tilde \phi) &=& \left.2e^Q(C - w)\right|_{\phi=\phi(\tilde \phi)}\,,\quad \extd \tilde \phi = \extd \phi e^Q\,,\\\label{eq:consquant}
 C &=& e^{Q(\phi)} Y + w(\phi)\,.
\eeqa
Here $f$ is a free function on two-dimensional space-time. By using $f$ and $\tilde \phi$ directly as coordinates the line element \eqref{eq:FOGlineelem} is found naturally in Eddington-Finkelstein gauge, although no diffeomorphism gauge fixing was employed. The conserved quantity \eqref{eq:consquant}, which enters the Killing norm \eqref{eq:Killingnorm}, is exactly the nontrivial integral of motion found in the Katanaev-Volovich model \eqref{eq:KatanaevVolovich}. The solutions \eqref{eq:FOGlineelem} can have many Killing horizons, and in fact can be extended globally\cite{Klosch:1996fi,Klosch:1996qv,Kummer:1995qh,Katanaev:1996bh,Grumiller:2002nm}. As for the Katanaev-Volovich model the integrability extends to cases with special ``chiral'' matter\cite{Grumiller:2002nm}.

FOG also allows for a convenient reformulation as a so-called Poisson-Sigma-Model (PSM)\cite{Ikeda:1994fh,Schaller:1994uj,Schaller:1994es}, illuminating some of its structure more clearly: Grouping together the target-space coordinates $X^I = (\phi, X^a)$ and the gauge fields $A_I = (\om,e_a)$, FOG \eqref{eq:SFOG} can be written as
\eq{\Stext{PSM} = \int\limits_{\cM_2} \left[ \extd X^I \wedge A_I + \frac12 P^{IJ} A_J \wedge A_I\right]\ ,}{eq:PSM}
where $P^{IJ} = \{X^I,X^J\}$ is a Poisson tensor related to a Schouten-Nijenhuis bracket defined on the manifold. Since the Poisson tensor must have a vanishing Nijenhuis tensor with respect to this bracket,
\begin{equation}
\label{eq:nijenhuis}
  P^{IL}\partial _{L}P^{JK}+ \mbox{\it perm}\left( IJK\right) = 0\ ,
\end{equation}
the PSM action is invariant under the symmetries
\begin{align}
\label{eq:symtrans}
  \delta X^{I} &= P^{IJ} \ve _{J}\ , & \delta A_{I} &= -\mbox{d} \ve
  _{I}-\left( \partial _{I}P^{JK}\right) \ve _{K}\, A_{J}\ .
\end{align}
For \( P^{IJ} \) linear
in \( X^{I} \) the symmetries constitute a linear Lie algebra and
\eqref{eq:nijenhuis} reduces to the Jacobi identity for the structure
constants of a Lie group. Finally, we mention that the variation of $A_I$ and $X^I$ in \eqref{eq:PSM} yields the PSM
field equations
\begin{align}
\label{eq:gPSeom1}
  \extd X^I + P^{IJ} A_J &= 0\ ,&
  \extd A_I + \frac{1}{2} (\partial_I P^{JK}) A_K A_J &= 0\ .
\end{align}
All PSMs are in essence topological theories of gauge fields on a by itself dynamical target space. They also allow for the straightforward inclusion of other topological degrees of freedom (such as gauge fields in two dimensions). The central object in \eqref{eq:PSM} is the Poisson tensor $P^{IJ}$, which for bosonic FOG \eqref{eq:SFOG} reads
\beq\label{eq:PoissonFOG}
P^{\phi\pm} = \pm X^\pm\,, \quad P^{+-} = X^+X^-U(\phi) + V(\phi)\,,\quad P^{IJ}=-P^{JI}\,.
\eeq
The PSM formulation of FOG is very useful in the context of supergravity, see Sect.~\ref{sect:sugra}. Since the Poisson tensor \eqref{eq:PoissonFOG} has odd dimension it cannot have full rank. Therefore there appears at least one ``Casimir function'' defined by the condition
\begin{equation}
\label{eq:casimir}
  \{ X^I, C \} = P^{IJ} \frac{\partial C}{\partial X^J} = 0\ .
\end{equation} 
The conserved quantities are thus central elements of the algebra of target space coordinates under the Schouten-Nijenhuis bracket. It is straightforward to show the conservedness by using the equations of motion \eqref{eq:gPSeom1}, as well as to reproduce the form of \eqref{eq:consquant}.

\subsection{Exact Path Integral Quantization}\label{sect:exactpathintegral}

In the years 1997-1999 Wolfgang Kummer, Herbert Liebl 
and Dimitri Vassilevich\cite{Kummer:1997hy,Kummer:1997jj,Kummer:1998zs} found that an 
exact path integral quantization is possible for the first order formulation \eqref{eq:SFOG}, \eqref{eq:SGDT}.
Even more, if \eqref{eq:SFOG} is coupled to matter\cite{Kummer:1997hy,Kummer:1997jj,Kummer:1998zs,Fischer:2001vz,Grumiller:2001rg,Bergamin:2004aw,Grumiller:2006ja,Meyer:2005fz,Meyer:2006vha,Meyer:2006xm}, the geometric sector $(\om,e^a,X^a,\phi)$ can still be 
integrated out exactly, yielding a nonlocal and nonpolynomial effective action for the matter fields, 
which then can be treated perturbatively. In this section, we will shortly review the most 
important steps in the path integral quantization, more details are explained in the context of supergravity in Sect.~\ref{sect:sugra} or can be found e.g.\ in the review \refcite{Grumiller:2002nm}. 

The goal is to evaluate the path integral
\begin{equation}
\label{eq:genfunct}
 \mathcal Z = \int \mathcal D(\Phi, \phi, X^a, \omega_\mu, e_\mu^a) e^{i\Stext{FOG}[\phi,X^a,\om,e^a]+i\Stext{mat}[\phi,e^a,\Phi]}
\end{equation}
where $\Phi$ denote all kinds of matter fields, which should not couple directly to the spin connection if the equivalence with generalized dilaton theories \eqref{eq:SGDT} shall not be lost. Since the model can be formulated as a nonlinear gauge theory with diffeomorphisms and local Lorentz invariance as symmetries, the first step is to construct an appropriate gauge fixed action. To this end, one analyzes the constraints in the theory, by first noting that \eqref{eq:SFOG} furnishes a natural symplectic structure with canonical variables\footnote{Note that these conventions accord with most literature on supergravity\cite{Bergamin:2002ju,Bergamin:2004us} but differ from those in Ref.~\refcite{Grumiller:2002nm}, where the components of the gauge fields were chosen as momenta rather than canonical coordinates.} $q^I = (\phi,X^a)$, $p_I = (\om_1, e_1^a)$, and $\bar p_I = (\om_0, e_0^a)$. The second set of momenta $\bar p_I$ does not appear with time derivatives in the Lagrangian and thus its  conjugate coordinates are constrained to zero, $\bar q^I \approx 0$, which constitutes three primary first class constraints. These constraints give rise to three secondary first class constraints $G_I = \{\bar p_I,\cH \}$. If the matter extension yields additional second class constraints, as e.g.\ in the case of fermions\cite{Grumiller:2006ja}, the Poisson bracket should be replaced by a the Dirac bracket. The three constraints $G_I$ form a closed nonlinear algebra 
\begin{equation}
\label{eq:constralg}
\{G_I, G'_J\} = G^K f_K{}^{IJ} \delta(x-x')\,,
\end{equation}
with structure functions $f_{K}{}^{IJ}$. A Virasoro-like algebra closing on derivatives of $\delta$-functions typical for 2D gravity models can be recovered from linear combinations of the $G^I$\cite{Katanaev:1994qf,Kummer:1994ur}. As expected for a theory of gravity the Hamiltonian density vanishes on the constraint surface,
\begin{equation}\label{geomham}
 \cH = \dot q^I p_I - L = - G^I \bar p_I\ .
\end{equation}
From this knowledge it is now straightforward to construct the gauge fixed action with ghosts, using the BVF formalism \cite{Fradkin:1975cq,Fradkin:1978xi,Batalin:1977pb}: Introducing one pair of ghosts and antighosts for each secondary first class  constraint, $(c_I,p_c^I)$,\footnote{Strictly speaking it would also be necessary to introduce (anti)ghosts for the primary constraints. The procedure is straightforward, as described in chapter 7 of Ref.~\refcite{Grumiller:2002nm}, and does not yield much new insight.} one finds the nilpotent BRST charge
\begin{equation}
  \Omega =  G^I c_I + \frac{1}{2} p_c^K f_K{}^{IJ} c_J c_I\,, \qquad \{\Omega,\Omega\} = 0\,. \label{eq:BRSTcharge}
\end{equation}
having the same structure as in ordinary Yang-Mills theory, albeit the field-dependence in $f_K{}^{IJ}$. At this point it is important to follow Wolfgang Kummer's concept of temporal gauges and to fix e.g.\ $(\om_0, e_0^-, e_0^+) = (0,1,0)$, which again establishes Eddington-Finkelstein gauge for the metric. This can be implemented by a simple multiplier gauge $\Psi = p_c^+$\cite{Grumiller:2001ea}, another possibility was used in Ref.~\refcite{Kummer:1997hy}.
Now the gauge fixed Hamiltonian $\cH_{gf} = \{\Om,\Psi\}$, via Legendre transformation\footnote{Of course the Legendre transform also has to be done w.r.t. possible matter canonical variables present. We tacitly ignored this here, as we do not want to specify the matter content yet, but describe the overall structure.} yields the gauge fixed Lagrangian, 
\eq{\cL_{gf} = \dot{\bar q}^I \bar p_I +  \dot q^I p_I + G^+ + p_c^K (\partial_0 \delta_K^L + f_K{}^{+L}) c_L\,.}{eq:gaugefixedL}

The path integral \eqref{eq:genfunct} can now be evaluated as follows: First all (anti)ghosts are integrated out yielding the Faddeev-Popov determinant $\Det \De_{FP} = \Det (\partial_0 \delta_K^L + f_K{}^{+L})$. Now an important observation is made: Eq.~\eqref{eq:gaugefixedL} only depends linearly on $p_I$, while $\Det \De_{FP}$ is independent thereof. After eventual integration of matter momenta the geometric fields $p_I$ thus can be integrated. This generates three ``functional $\delta$ functions'', whose arguments contain parts of the equations of motion and imply in the matterless case ($j^I$ are sources for the $p_I$) 
\begin{gather}
\label{eq:delta1}
 0 = \partial_0 \phi - X^+ - j^\phi\,, \qquad 0 = \partial_0 X^+ - j^+\,,\\
\label{eq:delta3}
 0 = (\partial_0 + X^+ U(\phi)) X^- + V(\phi) - j^-\,,
\end{gather}
The final integration of $X^I = q^I$ sets these fields to the formal solution of the above equations given in terms of Green functions for $\partial_0$ acting on the sources. Note that these solutions depend nonlocally on the sources, and in particular the solution of \eqref{eq:delta3} is also nonpolynomial. In defining the Green functions the asymptotic values of $X^I$ are fixed, which also fixes the quasilocal mass. For this reason, the path integral as it is yields local physics. It is a question worth investigating whether in the path integral formalism an ``integration over masses'' is possible --- see Sect.~\ref{sect:boundary} for some further comments on this topic. Finally, it should be mentioned that the integration over $q^I$ cancels exactly the Faddeev-Popov determinant.

In the absence of matter fields the story ends here since the path integral is fully evaluated. If an integration over matter fields is left, the so-far obtained effective action is a complicated, nonlocal and nonpolynomial expression in the matter fields. Still, it is possible to derive in a systematic way non-local vertices of the interaction of matter with the quantized gravitational background. Therefrom, higher loops\cite{Kummer:1997jj} and scattering processes can be calculated\cite{Fischer:2001vz}. For a real scalar field unitarity, i.e.\ absence of information loss, and CPT invariance of the tree-level four-particle S-matrix element was found. The specific heat of the Witten black hole was also shown to be positive\cite{Grumiller:2002dm} once loop corrections are taken into account.

A phenomenon worth mentioning is the ``virtual black hole (VBH)''\cite{Grumiller:2002dm,Grumiller:2000ah,Grumiller:2001rg,Grumiller:2004yq}: Some of the nonlocal interaction geometries resemble black holes, although being off-shell entities. The curvature scalar of a VBH has a $\de$-peak at some point and is discontinuous up to that peak (see the Penrose diagram in Fig.~2 of ref.~\refcite{Grumiller:2006rc}, an effective line element can also be found there). Although this interpretation is a feature of the chosen gauge, in the calculation of (gauge invariant) S-matrix elements one has to integrate over all such VBHs, which leads to the idea that nonperturbative and nonlocal excitations such as VBHs might also play an important role in higher-dimensional quantum gravity.

%% file: part2.tex
\section{Two-Dimensional Dilaton Supergravity}\label{sect:sugra}
In the following we will leave the field of bosonic dilaton gravity, since many results thereof have been summarized in various reviews\cite{Grumiller:2002nm,Grumiller:2006rc}. Instead we will concentrate on dilaton supergravity in two dimensions, a field in which Wolfgang Kummer made substantial contributions as well but which is not covered exhaustively by the existing reviews.

Of course, dilaton supergravity in two dimensions existed long before Wolfgang Kummer entered the field. Many early attempts were based on superspace techniques\cite{Howe:1979ia}, which led to a generalized dilaton supergravity action of the form\cite{Park:1993sd}
\begin{equation}
  \label{eq:ps1}
  \mathcal S = \int\! \diff{^2x} \diff^2 \theta E \bigl( \Phi S - \frac{1}{4}
  U(\Phi) D^\alpha \Phi D_\alpha \Phi + \frac12 u(\Phi) \bigl)\ ,
\end{equation}
where $S$ and $\Phi$ are the supergravity and dilaton superfields, resp, and $U(\Phi)$ and $u(\Phi)$ two dilaton dependent functions (potentials) defining the model.\footnote{In Ref.~\refcite{Park:1993sd} a third potential $J(\Phi) S$ was introduced in the first term. If $J(\Phi)$ is not invertible, the analysis performed here applies locally, only. However, such models are of limited interest.\cite{Strobl:1999wv,Grumiller:2003dh}} Despite its successes and advantages,\cite{Park:1993sd,Bilal:1993ss,Nojiri:1993sx} in particular the straightforward way to couple additional gauge or matter fields, the superspace formulation shares its drawbacks with the purely bosonic second order action: It does not include bosonic torsion, exact solutions are not easy to obtain although the matterless model is integrable, and a nonperturbative quantization is cumbersome. Some attempts to relax the standard condition of vanishing bosonic torsion led to a mathematically complex formalism\cite{Ertl:1998ib,Ertl:2001sj} and thus were not pursued further.

\subsection{2D Dilaton Supergravity from Graded PSMs}
In this situation it appeared promising to extend FOG \eqref{eq:SFOG} to dilaton supergravity. Already in Refs.\ \refcite{Rivelles:1994xs,Cangemi:1993mj} gauge theoretic methods were used to find supersymmetric extensions of some dilaton gravity models. However, this approach is limited to models representable as a linear gauge theory, in particular the Jackiw-Teitelboim model.\cite{Teitelboim:1983ux,jackiw:1984} Its extension to nonlinear gauge theories showed that generalized dilaton supergravity can be treated in this framework.\cite{Ikeda:1994dr,Ikeda:1994fh} This led to a first order action for dilaton supergravity using free differential algebras,\cite{Izquierdo:1998hg} which however is restricted to vanishing bosonic torsion. This model, which is classically equivalent to the action \eqref{eq:ps1} for $U=0$, then was shown to be a special case of a graded PSM.\cite{Strobl:1999zz}

Wolfgang Kummer, Martin Ertl and Thomas Strobl then showed\cite{Ertl:2000si} that the use of graded PSMs (gPSMs) provides a simpler and more systematic tool to find supergravity extensions of FOG. To this end one replaces the target space (Poisson manifold) of the action \eqref{eq:PSM} by a graded Poisson manifold.\footnote{Instead of a graded target space, supersymmetric PSMs can be obtained by grading the world-sheet manifold. While such models attracted interest in the context of string theory\cite{Bergamin:2004sk,Calvo:2005ww}, they are less useful to derive supergravity actions.} On this manifold we use the coordinates (scalar fields) $X^I(x) = (X^i(x), X^\alpha(x))$, where $X^i$ refers to bosonic (commuting) coordinates, while $X^\alpha$ are fermions (anti-commuting coordinates.) All equations \eqref{eq:PSM}--\eqref{eq:casimir} have been displayed in such a way that they hold for gPSMs as well, if one replaces the permutations in \eqref{eq:nijenhuis} by graded permutations and keeps in mind that $P^{IJ}$ is now graded anti-symmetric.

As in the non-supersymmetric case, not every gPSM describes a supergravity model, but certain additional structures are needed. Firstly this concerns the choice of target space and gauge fields. In the application to two-dimensional $N = (1,1)$ supergravity, the bosonic fields in \eqref{eq:SFOG} are complemented by two Majorana spinors, \( \psi _{\alpha } \) (``gravitino'') and \( \chi^{\alpha } \) (``dilatino''):
\begin{align}
\label{eq:MFSvariables}
  A_I & = (\omega, e_a, \psi_\alpha) & X^I & = (\phi, X^a, \chi^\alpha)
\end{align}
Secondly, local Lorentz invariance determines the $\phi$-components of the Poisson
tensor
\begin{align}
\label{eq:lorentzcov}
  P^{a \phi} &= X^b {\epsilon_b}^a\ , & P^{\alpha \phi} &= -\frac{1}{2}
  \chi^\beta {\gthree_\beta}^\alpha\ .
\end{align}
Finally it should be noted that the existence of a tanget space metric $\eta_{ab}$ is assumed as an extra structure.\footnote{Attempts to relax some of these conditions have been presented in Refs.\ \refcite{Strobl:2003kb,Adak:2007xc}.} Remarkably it was possible to solve the nonlinear Jacobi identity \eqref{eq:nijenhuis} explicitly with just these three assumptions.\cite{Ertl:2000si,Ertl:2001sj} Having defined the fermionic extension, all components of the graded Poisson tensor may be written as a systematic expansion in these fields, restricted by local Lorentz invariance encoded in the $\ve_\phi$ component in \eqref{eq:symtrans}. As an example $P^{ab}$ must be of the form
$P^{ab} = \cV(\phi,Y) + \chi^2 v_2(\phi,Y)$,
where $\cV(\phi,Y)$ is the bosonic potential \eqref{eq:GDTpotential} of the model. Then it is a straightforward but tedious calculation to solve the condition \eqref{eq:nijenhuis} order by order in the fermionic fields. Though a purely algebraic solution was found\cite{Ertl:2000si,Ertl:2001sj}, this result was not yet satisfactory, as it depends besides the bosonic potential $\cV(\phi,Y)$ on five arbitrary Lorentz covariant functions. However, many choices of these five five functions impose new singularities and obstructions in the variables $\phi$ and $Y$ in points where the bosonic potential remains regular. In extreme cases this prohibits any supersymmetrization at all.

From this result it is obvious that not all Lorentz covariant gPSMs permit an interpretation as dilaton supergravity, but a suitable implementation of supersymmetry transformations is needed.\footnote{This procedure is similar to the choice of ``consistent'' supergravity in the standard formulation.\cite{Freedman:1976xh,Freedman:1976py,Deser:1976eh,Grimm:1978kp}} This can be achieved by restricting the fermionic extension $P^{\alpha\beta}$ to the form
\begin{equation}
\label{susytrafo}
 P^{\alpha \beta} = -2 i X^a \gamma_a^{\alpha \beta} + Z(\phi,Y,\chi^2) \gthree^{\alpha \beta}\ ,
\end{equation}
where the first term generates supersymmetry transformations of the form $\{Q^\alpha,Q^\beta\} = -2i p^{\alpha\beta}$ in the commutator of two symmetry transformations \eqref{eq:symtrans}. The solution of the Jacobi identities now proved a unique class of $N=(1,1)$ dilaton supergravity models with Poisson tensor\footnote{Notice that the potential $U(\phi)$, which determines the torsion part of the spin connection, in most of the literature on supergravity is denoted by $Z(\phi)$.}
\begin{align}
\label{eq:mostgensup}
  P^{ab} &= \biggl( V + Y U - \frac{1}{2} \chi^2 \Bigl( \frac{VU + V'}{2u} +
  \frac{2 V^2}{u^3} \Bigr) \biggr) \epsilon^{ab}\ , \\
  P^{\alpha b} &= \frac{U}{4} X^a
    {(\chi \gamma_a \gamma^b \gthree)}^\alpha + \frac{i V}{u}
  (\chi \gamma^b)^\alpha\ , \\
\label{eq:mostgensuplast}
  P^{\alpha \beta} &= -2 i X^c \gamma_c^{\alpha \beta} + \bigl( u +
  \frac{U}{8} \chi^2 \bigr) \gthree^{\alpha \beta}\ ,
\end{align}
whereby $u(\phi)$ acts as a prepotential for the bosonic potential $V(\phi)$
\begin{equation}
\label{eq:finalpot}
V\left( \phi \right) =- \frac{1}{8} \bigl(( u^{2})' + u^{2} U\left( \phi
\right) \bigr) \ .
\end{equation}
First, this result was established from symmetry arguments\cite{Bergamin:2002ju} by constructing a deformed superconformal algebra as previously done for FOG\cite{Katanaev:1994qf}. Then it turned out that the supergravity action from \eqref{eq:lorentzcov} and \eqref{eq:mostgensup}--\eqref{eq:mostgensuplast},
\begin{multline}
  \label{eq:mostgenaction}
  \Stext{} = \int_{\mathcal{M}} \bigl( \phi \diff \omega + X^a D e_a + \chi^\alpha D
  \psi_\alpha + \epsilon \biggl( V + Y U - \frac{1}{2} \chi^2 \Bigl( \frac{VU + V'}{2u} +
  \frac{2 V^2}{u^3} \Bigr) \biggr) \\
   + \frac{U}{4} X^a
    (\chi \gamma_a \gamma^b e_b \gthree \psi) + \frac{i V}{u}
  (\chi \gamma^a e_a \psi) \\
   + i X^a (\psi \gamma_a \psi) - \frac{1}{2} \bigl( u +
  \frac{U}{8} \chi^2 \bigr) (\psi \gamma_3 \psi) \bigr)\ ,
\end{multline}
after elimination of the auxiliary fields $X^a$ and $\omega$ is equivalent to the action \eqref{eq:ps1} after integrating out superspace.\cite{Bergamin:2003am} As in FOG there exists at least one Casimir function, which receives additional fermionic contributions:
\begin{equation}
  \label{eq:genbosc}
  C = e^{Q}\bigl(Y - \frac{1}{8}
  u^2 + \frac{1}{16} \chi^2 (u' + \frac{1}{2} u U) \bigr)\ .
\end{equation}
The Casimir function plays a crucial role in the discussion of BPS states.\cite{Bergamin:2003mh} From the symmetry transformation of the dilatino in Eq.~\eqref{eq:symtrans} it follows that the fermionic part of the Poisson tensor for BPS solutions does not have full rank. For purely bosonic states this implies $\Det P^{\alpha\beta} = 8 Y - u^2 = 0$ and thus $C=0$. Furthermore, an additional fermionic Casimir function exists and it can be checked\cite{Ertl:2000si,Bergamin:2003am} that the quantity
\begin{equation}
 \tilde{c} = e^{\frac{1}{2} Q} \sqrt{|X^{++}|}\left(\chi^- - \frac{\chi^+}{2\sqrt{2}X^{++}}\right)
\end{equation}
for $C=0$ is a Casimir function as defined in \eqref{eq:casimir}, which corresponds to the unbroken supersymmetry of the BPS state.\cite{Bergamin:2003mh}
Besides the purely bosonic states there also exist BPS states with non-vanishing fermionic fields, which however must have a trivial bosonic background.\cite{Bergamin:2003mh} Nonetheless, some of these states exhibit the interesting feature that $\chi^2 (u' + \frac{1}{2} u U)\neq 0$ and thus the bosonic part of the Casimir vanishes, while its fermionic extension (its soul) is non-zero.

Also the complete classical solution now is straightforward to obtain.\cite{Ertl:2000si,Ertl:2001sj,Bergamin:2003am} The bosonic line element again may be written in the form \eqref{eq:FOGlineelem}. From this result the extremality of BPS Killing horizons\cite{Gibbons:1982fy,Tod:1983pm} is immediate since by virtue of the constraint \eqref{eq:finalpot} the conformally invariant potential acquires the form $w(\phi) = -e^Qu^2/8$. Thus, the Killing norm $\xi$ in Eq.\ \eqref{eq:Killingnorm} for BPS solutions is positive semi-definite, an eventual horizon must be extremal and represents a ground state.\cite{Bergamin:2003mh}

Coupling of additional fields (e.g.\ gauge fields or matter fields) is straightforward for FOG, but in supergravity this task becomes considerably more complicated since a correct supersymmetry transformation of the new fields must be ensured. Since it is known that all genuine gPSM supergravity models are classically equivalent to superspace models, the coupling of matter fields can be inferred therefrom. As an example, it has been demonstrated in Ref.\ \refcite{Bergamin:2003mh} that conformal matter fields can be coupled non-minimally to the general action \eqref{eq:mostgenaction}. The matter action for a chiral multiplet with real scalar field $f$ and  Majorana fermion $\lambda_\alpha$ and with coupling function $K(\phi)$ for second order supergravity is obtained from standard superspace techniques as
\begin{multline}
  \label{eq:cmosh}
    \Stext{matter} = \int\! \diff{^2 x} e \Biggl[ K \Bigl(\frac{1}{2}
    (\partial^m f \partial_m f + i \lambda \gamma^m \partial_m \lambda)
    + i (\psi_n \gamma^m \gamma^n \lambda) \partial_m f\\
    + \frac{1}{4} (\psi^n
    \gamma_m \gamma_n \psi^m) \lambda^2 \Bigr)  +
    \frac{u}{8} K' \lambda^2 - \frac{1}{4} K' 
    (\chi \gthree \gamma^m \lambda) \partial_m f \\
    - \frac{1}{32} \Bigl(K''  - \frac{1}{2} \frac{\bigl[K' \bigr]^2}{K}\Bigr) \chi^2 \lambda^2 \Biggr]\ .
\end{multline}
Without further changes this action provides the correct extension of genuine gPSM supergravity by a conformal matter field. Still, the supersymmetry transformations of some of the gPSM fields acquire additional terms from the matter action.\cite{Bergamin:2003mh}

The gPSM approach to two-dimensional dilaton supergravity is not restricted to the above example of $N=(1,1)$ supergravity. As an example the gPSM version of the model of Ref.~\refcite{Nelson:1993vm} has been presented in Refs.~\refcite{Bergamin:2004sr,Bergamin:2004na}. This model deals with chiral or twisted-chiral $N=(2,2)$ supergravity. Thus besides a pair of complex Dirac dilatini/gravitini an additional graviphoton gauge field is needed in the supergravity multiplet, while the (twisted-)chiral dilaton multiplet is complemented by an additional Lorentz scalar $\pi$:
\begin{align}
    \label{eq:psm9}
    X^I &= (\phi, \pi, X^a, \chi^\alpha, \bar{\chi}^\alpha)\ , & A_I &= (\omega,
    B, e_a, \psi_\alpha, \bar{\psi}_\alpha)\ .
\end{align}
Again symmetry constraints fix certain components of the Poisson tensor. Local Lorentz invariance and supersymmetry are implemented analogously to \eqref{eq:lorentzcov} and \eqref{susytrafo} and supersymmetry is encoded as $P^{\alpha \bar{\beta}} = - 2i X^c \gamma_c^{\alpha \beta} + \mbox{terms
    $\propto \gthree$}$. The additional $B$ gauge symmetry imposes (here for chiral supergravity)
\begin{align}
     \label{eq:psm11}
    P^{a \pi} &= 0 , &  P^{\alpha
    \pi} &= - \frac{i}{2} \chi^\beta \gthree{}_\beta{}^\alpha\ , &
    P^{\bar{\alpha} \pi} = \frac{i}{2} \bar{\chi}^\beta
    \gthree{}_\beta{}^\alpha\ .
\end{align}
Starting with these restrictions it is again possible to solve the non-linear Jacobi identity \eqref{eq:nijenhuis} and to derive the complete action of extended supergravity.\cite{Bergamin:2004sr} Again, once reformulated as a gPSM the model can be solved explicitly, whereby some subtleties arise as a consequence of the grading of the Poisson manifold.\cite{Bergamin:2004na} The model is found to have at least two Casimir functions, one being the $N=(2,2)$ extension of \eqref{eq:consquant}, the second one is the charge with respect to the graviphoton field. This new conserved quantity diverges in the limit $C\rightarrow 0$ unless all fermionic contributions vanish. Nevertheless, after integrating all equations of motion it is found that there exist well-defined solutions at $C=0$ with non-vanishing fermion fields. However, the remaining conserved quantity (the charge with respect to the graviphoton) can no longer be expressed entirely in terms of the target space variables, i.e.\ for those special solutions this conserved quantity is not a Casimir function of the graded Poisson manifold. This is a unique feature of graded PSMs, since for purely bosonic PSMs the existence of Casimir-Darboux coordinates excludes such a behavior.

\subsection{Quantization of 2D Supergravity}
Having formulated dilaton supergravity as a gPSM it is self-evident to try to extend the nonperturbative quantization outlined in Sect.~\ref{sect:exactpathintegral} to dilaton supergravity.\cite{Bergamin:2004us,Bergamin:2004aw}

\subsubsection{Quantization Without Matter}
\label{sec:NMQM}
Let us first consider the dilaton supergravity action \eqref{eq:mostgenaction} without matter couplings. As in Sect.~\ref{sect:exactpathintegral} the canonical coordinates are chosen as $q^I = X^I$, $p_I = A_{1I}$, $\bar p_I = A_{0I}$ and $\bar q^I \approx 0$ are the primary first class constraints. In supergravity these coordinates obey the graded Poisson bracket\footnote{This bracket should not be confused with the Schouten bracket, associated to the Poisson tensor $P^{IJ}$ in \eqref{eq:PSM}.}
\begin{equation}
\{ q^I , p'_J \} = (-1)^{I\cdot J+1} \{p'^J, q_I\} = (-1)^{I}\delta^J_I \delta(x-x')\ .
\end{equation}
The secondary constraints
\begin{equation}
\label{geomG}
G^I = \partial_1 q^I + P^{IJ} p_J\ ,
\end{equation}
despite the new fermionic constraints formally obey the same constraint algebra \eqref{eq:constralg} as in the purely bosonic case. The structure functions in the matterless case are simply $f_K{}^{IJ} = -\partial_K P^{IJ}$ with $P^{IJ}$ as given in \eqref{eq:lorentzcov} and \eqref{eq:mostgensup}--\eqref{eq:mostgensuplast}, and with fermionic derivatives being left-derivatives. The ensuing Hamiltonian \eqref{geomham} again is quantized by introducing ghosts, which now obey the graded commutation rules
  $[c_I,p_c^J ] = - (-1)^{(I+1)(J+1)} [ p_c^J, c_I ] =  \delta_I^J$.
This yields an additional sign in the BRST charge\cite{Bergamin:2004us}
\begin{equation}
\label{eq:omegadef}
  \Omega = G^I c_I + \frac{1}{2} (-1)^I p_c^K f_K{}^{IJ} c_J c_I\ .
\end{equation}
The BRST charge \eqref{eq:omegadef} is found to be
nilpotent as a consequence of the graded Jacobi identity of $P^{IJ}$ \eqref{eq:nijenhuis}. This characteristic does not depend on the specific form of the Poisson tensor as found in  \eqref{eq:mostgensup}--\eqref{eq:mostgensuplast} and thus also applies to graded dilaton models not related to the supergravity action \eqref{eq:ps1}. Also, any gPSM gravity model without matter is free of ordering problems although the constraints are nonlinear in the fields.\cite{Bergamin:2004us}

Imposing again a multiplier gauge $\bar p_I = a_I$ with\footnote{Notice that the elements of the spin-tensor decomposition are related to lightcone indices as $v^{++} = i v^{\oplus}$, $v^{--} = -i v^{\ominus}$, cf.\ Eq.~\eqref{lightcone}. Thus $\bar p_{++}$ is imaginary.} $a_I = -i \delta_I^{++}$ one arrives at the gauge fixed Lagrangian
\begin{equation}
\label{eq:quant11.2}
  L^0_{\mbox{\tiny g.f.}} = \dot{q}^I p_I + p_c^I \dot{c}_I - i P^{++|J} p_J \ .
\end{equation}
The evaluation of the path integral now follows exactly the same steps as in the purely bosonic case.
Having integrated out all ghosts the effective Lagrangian including sources for $p_I$ and $q^I$ becomes
\begin{equation}
  \label{eq:quant16}
  L^0_{\mbox{\tiny eff}} = \dot{q}^I p_I - i P^{++|J} p_J + q^I j_{qI} + j_p^I
    p_I
\end{equation}
which again is linear in all $p_I$. Integration of the momenta thus yields five functional $\delta$ functions, whose arguments imply (for the Poisson
tensor \eqref{eq:lorentzcov}, \eqref{eq:mostgensup}-\eqref{eq:mostgensuplast})
\begin{gather}
  \dot{q}^\phi = - i q^{++} - j_p^\phi\ , \label{eq:quant17.1}\qquad \dot{q}^{++} = - j_p^{++}\ , \\
  \dot{q}^{--} = i \Bigl( e^{-Q} w' + q^{++} q^{--} U + \frac{1}{2 \sqrt{2}} q^- q^+
  e^{-Q/2}\bigl( \sqrt{-w} \bigr)''\Bigr)- j_p^{--}\ ,
  \label{eq:quant17.3}\\
  \dot{q}^{+} = - j_p^{+}\ , \qquad 
  \dot{q}^{-} = - i \bigl(e^{-Q/2}(\sqrt{-w})' q^+ - \frac{1}{2} U q^{++}
  q^- \bigr) - j_p^{-}\ . \label{eq:quant17.5}
\end{gather}
These functional $\delta$ functions may be used to evaluate the remaining $q^I$ integration. As expected, this exactly cancels the super-determinant from the integration of the ghosts. The final generating functional becomes
\begin{equation}
  \label{eq:quant19}
  \mathcal{W}[j_p^I, j_{qI}] = \exp{iL^0_{\rm eff}}\,, \quad L^0_{\mbox{\tiny eff}}=\int\! \diff^2x \left(B^I j_{qI} + \tilde{L}^0 (j_p^I, B^I)\right) \ ,
\end{equation}
where $B^I$ are the solutions of Eqs.\ \eqref{eq:quant17.1}--\eqref{eq:quant17.5} and $\tilde{L}^0$ are the so-called ``ambiguous terms''.\cite{Haider:1994cw,Kummer:1997hy,Grumiller:2002nm} An expression of this
type is generated by the integration constants $G_I(x^1)$ from the term $\int\! \diff
  x^0\int\!\diff y^0 (\partial_0^{-1}A^I) j_{qI}$.

From this result it is straightforward to calculate the quantum effective action
\begin{equation}
  \label{eq:745}
  \Gamma\left(\langle q^I\rangle,\langle p_I\rangle\right):=L^0_{\mbox{\tiny eff}}\left(j_{qI},j_p^I\right)-\int\! \diff^2x\left(\langle q^I\rangle j_{qI}+j_p^I\langle p_I\rangle\right)\ ,
\end{equation}
in terms of the mean fields
\begin{equation}
  \label{eq:668}
  \langle p_I\rangle:=\left. \frac{\stackrel{\rightarrow}{\delta}}{\delta j_p^I} L^0_{\mbox{\tiny eff}}\right|_{j=0}\ ,\quad\langle q^I\rangle:=\left. L^0_{\mbox{\tiny eff}} \frac{\stackrel{\leftarrow}{\delta}}{\delta j_{qI}} \right|_{j=0} = \left. B^I\right|_{j_p=0}\ .
\end{equation}
By re-expressing all sources in terms of the (classical) target space coordinates it is found\cite{Bergamin:2004us} that---up to boundary terms---the quantum effective action is nothing but the gauge fixed classical action. Thus similar to FOG\cite{Kummer:1997hy} two-dimensional dilaton supergravity exhibits local quantum triviality. Still, the quantization procedure is not completely trivial as the generating functional \eqref{eq:quant19} is essentially non-local.

\subsubsection{Quantization Including Matter Fields}
Local quantum triviality provides an important consistency check, the main purpose of the nonperturbative quantization procedure is the straightforward way to couple matter fields perturbatively to the fully quantized geometric background.\cite{Kummer:1997jj,Kummer:1998zs,Grumiller:2001ea,Grumiller:2002nm} While an immediate extension to supergravity could be expected for the matterless theory on general grounds, this came as a surprise for the theory including matter fields:\cite{Bergamin:2004us,Bergamin:2004aw} The matter action \eqref{eq:cmosh} is obtained from the superspace action by a non-trivial integration of auxiliary fields and exactly this step, in the second order formalism, generates quartic ghost terms, which would spoil the nonperturbative quantization procedure. In the first order formulation this does not happen, which makes the full integration over geometry possible in the first place.

From the matter action \eqref{eq:cmosh} together with the matter fields and momenta, $\matq = f$, $\matp = \partial L_{(m)}/ \partial \dot{\matq}$ and $\matq^\alpha = \lambda^\alpha$, $\matp_\alpha = \partial L_{(m)}/ \partial \dot{\matq}^\alpha$,
%
%
the Hamiltonian density follows as $H_{(m)} = \dot{\matq} \matp + \dot{\matq}^+ \matp_+ + \dot{\matq}^- \matp_- - L_{(m)}$. The total Hamiltonian density is the sum of this contribution and
\eqref{geomham}. For the Poisson bracket of two matter field monomials one finds
 $\{ \matq, \matp' \} = \delta(x - x')$ and  $ \{ \matq^\alpha, \matp'_\beta\} = -
  \delta^\alpha_\beta  \delta(x - x')$.
We do not provide the explicit form of the matter
Hamiltonian, as it can again be written in terms of secondary constraints, $H = G^I \bar{p}_I$. The constraints itself divide into matter and geometry, $ G^I = G^I_{(g)} + G^I_{(m)}$, where $G^I_{(g)}$ has been derived in \eqref{geomG} and the matter part reads ($\partial=\partial_1$):
{\allowdisplaybreaks
\begin{align}
\label{eq:G++m}
\begin{split}
  G^{++}_{(m)} &= - \frac{K}{4 p_{++}} (\partial \matq - \frac{1}{K} \matp)^2 +
  i (\partial \matq - \frac{1}{K} \matp) (\frac{K}{p_{++}} p_+
  \matq^+  - \frac{K'}{4 \sqrt{2}} \frac{p_{--}}{p_{++}} q^- \matq^-)\\
   &\quad +  \frac{i K'}{4 \sqrt{2}}  (\partial \matq + \frac{1}{K}
   \matp) q^+ \matq^+  + \frac{K}{\sqrt{2}} \matq^+ \partial \matq^+ -
   \frac{K'}{2 \sqrt{2}} \frac{p_{--}}{p_{++}} p_+ q^- \matq^- \matq^+\\
   &\quad - p_{--} \Bigl( \frac{u K'}{4} - \frac{1}{8} (K'' -
   \frac{K'^2}{K}) q^- q^+ \Bigr) \matq^- \matq^+
\end{split}\,,
  \\
\label{eq:G--m}
\begin{split}
  G^{--}_{(m)} &=  \frac{K}{4 p_{--}} (\partial \matq + \frac{1}{K} \matp)^2 -
  i (\partial \matq + \frac{1}{K} \matp) (\frac{K}{p_{--}} p_-
  \matq^-  + \frac{K'}{4 \sqrt{2}} \frac{p_{++}}{p_{--}} q^+ \matq^+)\\
  &\quad +  \frac{i K'}{4 \sqrt{2}}  (\partial \matq - \frac{1}{K}
  \matp) q^- \matq^- - \frac{K}{\sqrt{2}} \matq^- \partial \matq^- +
   \frac{K'}{2 \sqrt{2}} \frac{p_{++}}{p_{--}} p_- q^+ \matq^- \matq^+\\
  &\quad + p_{++} \Bigl( \frac{u K'}{4} - \frac{1}{8} (K'' -
   \frac{K'^2}{K}) q^- q^+ \Bigr) \matq^- \matq^+
\end{split}\,,
  \\
\label{eq:G+m}
  G^+_{(m)} &= i K (\partial \matq - \frac{1}{K}\matp) \matq^+ -
  \frac{K'}{2 \sqrt{2}} p_{--} q^- \matq^- \matq^+ \,,\\
\label{eq:G-m}
  G^-_{(m)} &= -i K (\partial \matq + \frac{1}{K}\matp) \matq^- +
  \frac{K'}{2 \sqrt{2}} p_{++} q^+ \matq^- \matq^+\,.
\end{align}
}
As the kinetic term of the matter fermion $\lambda$ is first order only, this
part of the action leads to constraints as well. From $\matp_+ = - K p_{++} \matq^+/\sqrt{2}$ and $\matp_- = K p_{--} \matq^-/\sqrt{2}$,  the usual primary
second-class constraints are deduced:
\begin{align}
\label{eq:Psiconstr+}
   \Psi_+ &= \matp_+ + \frac{K}{\sqrt{2}} p_{++} \matq^+ \approx 0 \,,&
   \Psi_- &= \matp_- - \frac{K}{\sqrt{2}} p_{--} \matq^- \approx 0\,.
\end{align}
These second class constraints are treated by substituting the Poisson bracket
by the ``Dirac bracket'' \cite{Dirac:1996} $\{ f, g \}^* = \{ f, g \} - \{f, \Psi_\alpha\} C^{\alpha \beta} \{\Psi_\beta,
  g\}$, where $C^{\alpha \beta} C_{\beta \gamma} = \delta^\alpha_\gamma$ and $C_{\alpha \beta}
  = \{\Psi_\alpha, \Psi_\beta\}$.
Despite the complexity of these expressions it can be shown\cite{Bergamin:2004aw} that the secondary constraints with respect to the Dirac bracket obey an algebra of the type \eqref{eq:constralg}.
In the case of minimal coupling, $K(\phi) = 1$, the structure functions of the matterless theory \eqref{eq:constralg} are reproduced,\cite{Bergamin:2004us} while in the generic case additional matter contributions to the structure functions arise, explicit expression can be found in Ref.~\refcite{Bergamin:2004aw}.

The quantization follows the same steps as in the matterless case. As most important observation a lengthy calculation unravels that the homological perturbation theory again stops at first order,\cite{Bergamin:2004us,Bergamin:2004aw} with the BRST charge as given in Eq.~\eqref{eq:omegadef}.
Furthermore it is again found that despite the different non-linearities the theory does not exhibit ordering problems.\cite{Bergamin:2004us,Bergamin:2004aw} Thanks to this unexpectedly simple result we can pursue to formulate a path integral along the same lines as for the matterless case. In the same temporal gauge as used there the gauge fixed Lagrangian becomes
{\allowdisplaybreaks
\begin{multline}
  \label{eq:quant12}
  L_{\mbox{\tiny g.f.}} = \dot{q}^I p_I + \dot{\matq} \matp + \dot{\matq}^\alpha
  \matp_\alpha + p_c^I \dot{c}_I
  - i P^{++|J} p_J  - i (-1)^K p_c^I C_I{}^{++|K} c_K  \\
   + \frac{i}{4} \frac{K}{p_{++}} (\partial \matq - \frac{1}{K} \matp)^2 +
  (\partial \matq - \frac{1}{K} \matp) (\frac{K}{p_{++}} p_+
  \matq^+  - \frac{K'}{4 \sqrt{2}} \frac{p_{--}}{p_{++}} q^- \matq^-)\\
   +  \frac{K'}{4 \sqrt{2}}  (\partial \matq + \frac{1}{K}
   \matp) q^+ \matq^+ - \frac{i}{\sqrt{2}} K \matq^+ \partial \matq^+ +
   \frac{i}{2 \sqrt{2}} K' \frac{p_{--}}{p_{++}} p_+ q^- \matq^-
  \matq^+ +\\
  + i p_{--} \Bigl( \frac{u K'}{4} - \frac{1}{8} (K'' -
   \frac{K'^2}{K}) q^- q^+ \Bigr) \matq^- \matq^+\ .
\end{multline}
}
As in the matterless case the path integrals of $\bar{q}^I$, $\bar{p}_I$ are trivial and the ghosts just yield the super-determinant $\mbox{sdet}  {M_I}^J = \mbox{sdet} \bigl( {\delta_I}^J \partial_0 + i f_I{}^{++|J} \bigr)$.
The fermionic momenta $\matp_\alpha$ by means of the constraint
\eqref{eq:Psiconstr+} are integrated trivially as well, while this is possible for $\matp$ after a quadratic completion. The ensuing determinant can be absorbed by the re-definition of the path-integral measure of $\matq$ and $\matq^\alpha$ with correct superconformal properties.\cite{Rocek:1986iz,Lindstrom:1988fr} This yields the
effective matter Lagrangian
{\allowdisplaybreaks
\begin{multline}
  \label{eq:quant15}
  L_{(m)} =  i
  K p_{++} \dot{\matq}^2 - \frac{K}{\sqrt{2}} p_{++} \dot{\matq}^+ \matq^{+} + \frac{K}{\sqrt{2}} p_{--}
  \dot{\matq}^- \matq^{-} \\ + \dot{\matq} \bigl(K \partial \matq - 2 i K
  p_+ \matq^{+} + \frac{i}{2 \sqrt{2}} K' (p_{--} q^- \matq^- + p_{++}
  q^+ \matq^+)\bigr)\\
  + \frac{K'}{2 \sqrt{2}} \partial \matq q^+ \matq^+ 
   - \frac{i}{\sqrt{2}} K \matq^+ \partial
  \matq^+\\ + i p_{--} \Bigl( \frac{u K'}{4} - \frac{1}{8} (K'' - \frac{1}{2}
   \frac{K'^2}{K}) q^- q^+ \Bigr) \matq^- \matq^+\ .
\end{multline}
}
Since this expression is again linear in $p_I$ all geometric
variables can be integrated out and one is left with the integration
of the matter variables, which must be treated
perturbatively. It should be noted that the above Lagrangian explicitly depends on the prepotential $u$ as a consequence of the elimination of
(superspace) auxiliary fields. This is different than in all bosonic models, where a strict separation of the potentials appearing in the geometric part ($V$ and $U$) and the one of the matter extension, $K$, occurs.

Having performed the remaining integrals of all geometric quantities one is left with a path integral in the matter fields $\matq$ and $\matq^\alpha$:
\begin{multline}
 \mathcal{W}[\mathcal{J}] =  \int \mathcal{D}(\matq,\matq^\alpha) \exp\biggr[i \int\! \extd^2x \Bigl( K(\dot \matq \partial \matq - \frac{i}{\sqrt{2}} \matq^+ \partial \matq^+)\\ + \frac{K'}{2\sqrt{2}} \partial \matq q^+ \partial \matq^+ + B^I j_{qI} + \tilde{L} (j_p^I, B^I) + \matq J + \matq^\alpha J_\alpha \Bigr)\biggl]\,.
\end{multline}
Here, the $B^I$ are solutions to the functional $\delta$ functions
\begin{gather}
\displaybreak[0]
  \dot{q}^\phi = q_{(g)}^{\phi}\ , \label{eq:quant21.1} \qquad
  \dot{q}^{++} = q_{(g)}^{++} - i K \dot{\matq}^2 + \frac{K}{\sqrt{2}}\dot{\matq}^+
  \matq^+ - i \frac{K'}{2 \sqrt{2}} \dot \matq q^+ \matq^+\ , \\
  \label{eq:quant21.3}
   \begin{split}
    \dot{q}^{--} &= q_{(g)}^{--} - \frac{K}{\sqrt{2}} \dot{\matq}^- \matq^- - i \frac{K'}{2\sqrt{2}} \dot \matq q^- \matq^- \\
    &\quad -i \left(\frac{u K'}{4} - \frac{1}{8}\left(K'' - \frac{1}{2}\frac{{K'}^2}{K}\right) q^-q^+\right)\matq^- \matq^+
   \end{split}\\
  \dot{q}^{+} =  q_{(g)}^{+} - 2i K \dot{\matq} \matq^+\ , \qquad 
  \dot{q}^{-} =  q_{(g)}^{-}\ , \label{eq:quant21.5}
\end{gather}
with the $q^I_{(g)}$ being the right hand sides of Eqs.\ \eqref{eq:quant17.1}-\eqref{eq:quant17.5}.

\subsubsection{Four-Point Vertices}
From the matter Lagrangian \eqref{eq:quant15} and the differential equations \eqref{eq:quant21.1}--\eqref{eq:quant21.5} it is possible to derive the non-local vertices of matter to lowest order (tree level.) These results were presented in Ref.~\refcite{Bergamin:2004us} for minimal coupling, $K(\phi) = 1$, here we derive the more general result for non-minimal couplings.
In this calculation the concept of localized matter\cite{Kummer:1998zs,Grumiller:2000ah,Grumiller:2002dm} is used, here we follow the notation of Ref.~\refcite{Bergamin:2004us} and define
{\allowdisplaybreaks
\begin{align}
  \Phi_i(x) &= \frac{1}{2} \partial_i \matq(x) \partial_0 \matq(x) \Rightarrow a_i \delta^2(x-y)\ , \label{eq:quant23.1} \\
  \Psi_i^{\pm \pm} (x) &= \frac{1}{2} \partial_i \matq^{\pm}(x) \matq^{\pm}(x) \Rightarrow
  b^{\pm \pm}_i
  \delta^2(x-y)\ ,\\
  \Pi_i^{\pm}(x) &= \partial_i{\matq}(x) \matq^{\pm}(x) \Rightarrow c^\pm_i \delta^2(x-y)\ , \\
  \Lambda(x) &= \matq^- \matq^+(x) \Rightarrow e \delta^2(x-y)\ .
  \label{eq:quant23.3}
\end{align}
}
%
Notice that with our choice of gauge only $\Psi_1^{++}$ and $\Pi_1^+$, but no terms in $\Psi_1^{--}$ or $\Pi_1^-$ appear in the interaction. Furthermore, due to supersymmetry $a_0$ and $b_0^{++}$ only appear in the linear combination
\begin{equation}
 A_0 = 2i a_0 - \sqrt{2} b_0^{++}\ ,
\end{equation}
which thus will be used as abbreviation in the following.
Thanks to the local quantum triviality of the matterless theory, the lowest order vertices can be determined from the matter interaction terms in the gauge-fixed Lagrangian \eqref{eq:quant15} by solving to first order in localized matter the \emph{classical} equations of motion of the geometrical variables involved.\cite{Kummer:1998zs,Grumiller:2002nm,Grumiller:2002dm} To this end the
asymptotic integration constants must be chosen in a convenient way. Following
the calculations of the purely bosonic case \cite{Kummer:1998zs,Grumiller:2002dm}
$q^\phi(x^0 \rightarrow \infty) = x^0$, which implies $q^{++}(x^0 \rightarrow
\infty) = i$. In addition $p_{--}(x^0 \rightarrow \infty) = i e^Q$ may now be
imposed. Finally we have to fix the asymptotic value of the Casimir function
\eqref{eq:genbosc}, $C(x^0 \rightarrow \infty) = C_\infty$. Due to the matter
interactions $\extd C \neq 0$, but the conservation law receives contributions
from the matter fields as well \cite{Bergamin:2003mh}. Finally, considering the asymptotic values of the fermions $q^+$ and $q^-$ we follow Ref. \refcite{Bergamin:2004us} and set $q^+_\infty = q^-_\infty = 0$. This considerably reduces the complexity of the equations of motion since all contributions quadratic in the fermions vanish as they are second order in localized matter.

It turns out that the gauge fixed equations with the above choice of the asymptotic values can be solved explicitly to first order in localized matter. Therefore all non-local four-point vertices can be evaluated exactly to lowest order.
To economize writing of the subsequent non-local quantities we introduce the notations
\begin{align}
 \left[f g\right]_{x^0} &= f(x^0)g(x^0)\,, & \left[f g\right]_{x^0\pm y^0} &= \left[f g\right]_{x^0} \pm \left[f g\right]_{y^0}\ .
\end{align}
Furthermore, the abbreviation $h_{xy} = \theta(y^0-x^0) \delta(x^1 - y^1)$ is used. We do not repeat the complete solution here, as an example the Casimir function gets the new contributions
\begin{multline}
  \label{eq:quant25}
    C = - m_\infty + \Bigl[ i A_0 \Bigl(m_\infty [K]_{y^0} +
    [ K w ]_{y^0} \Bigr)\\ + \sqrt{2} i [e^Q K]_{y^0} b_0^{--} - \frac{1}{4}[e^Q u K]_{y^0} e \Bigr] h_{xy}\ .
\end{multline}
$m_\infty$ is the integration constant
of $q^{--} = i e^{-Q} m_\infty + \ldots$, which, however, turns out to be equivalent to the asymptotic
value $-C_\infty$. Notice that all contributions except this integration
constant are proportional to $h$ and thus to first order in localized matter all geometric variables may be
replaced by their asymptotic values in that equation. Of course, this is
equivalent to the statement that $C$ is constant in the absence of matter fields.

For the explicit expressions of the vertices one derives the relevant interaction terms from \eqref{eq:quant15} as
\begin{multline}
  \label{vertint}
  L_{(m)} =  (2i \Phi_0 - \sqrt{2} \Psi_0^{++)}) K p_{++} + \sqrt{2} K p_{--} \Psi_0^{--} + 2 K \Phi_1\\
  + (\frac{i}{2\sqrt{2}} K' p_{++} q^+ - 2i K p_+) \Psi_0^+ + \frac{i}{2\sqrt{2}} K' p_{--} q^- \Psi_0^- \\
  + \frac{K'}{2 \sqrt{2}} q^+ \Pi_1^+ + \sqrt{2} i K \Psi_1^{++} + \frac{i}{4} u K' p_{--} \Lambda\ .
\end{multline}
With the solution obtained one now finds that the vertices depend on seven different functions:
{\allowdisplaybreaks
\begin{gather}
\label{vertex1}
\begin{split}
 V_1(x,y) &= - K(x^0) K(y^0) \Biggl( 2 [w]_{x^0-y^0} - (x^0-y^0)\biggl ( [w']_{x^0+y^0}\\&\quad + \left[\frac{K'}{K}(w+m_\infty)\right]_{x^0+y^0}\biggr)\Biggr) h_{xy}\ ,
\end{split}  \\
\begin{split}
 V_2(x,y) &= -\sqrt{2} i K(x^0) K(y^0) \Biggl( \left[\sqrt{-w}'\right]_{x^0-y^0}\left[\sqrt{-w}\right]_{x^0-y^0}\\ &\quad- \frac{1}{2} \left[\frac{K'}{K}(w+m_\infty)\right]_{x^0+y^0}\Biggr) h_{xy}\ ,
\end{split} \\
 V_3(x,y) = 2i K(x^0) K'(y^0) |x^0-y^0| \delta(x^1-y^1)\ , \\
 V_4(x,y) = - \frac{i}{4} K(x^0)\left[e^Q u K'\right]'_{y^0} |x^0-y^0| \delta(x^1-y^1)\ , \\
 V_5(x,y) = \sqrt{2} K(x^0) \left[e^Q K'\right]'_{y^0} |x^0-y^0| \delta(x^1-y^1)\ , \\
  V_6(x,y) = -\frac{1}{\sqrt{2}} K(x^0) \left[\sqrt{-w}\right]_{x^0-y^0} \left[e^{\frac{Q}{2}} K'\right]_{y^0} (h_{xy}-h_{yx})\ ,\\
 \label{vertex7}
 V_7(x,y) = - \frac{i}{\sqrt{2}} K(x^0) K'(y^0) (h_{xy}-h_{yx})\ .
\end{gather}
}
The full interaction vertices from these functions are obtained as\cite{Grumiller:2002dm,Grumiller:2002nm,Bergamin:2004us}
\begin{equation}
 \mathcal V = \int_{x} \int_{y} \Xi_i(x) \Xi_j(y) V_{ij}(x,y)\ ,
\end{equation}
where $\Xi_i$ is one of the contributions from localized matter according to \eqref{eq:quant23.1}--\eqref{eq:quant23.3} and $V_{ij}$ is---up to eventual constants---the vertex function from \eqref{vertex1}--\eqref{vertex7} that describes the interaction between $\Xi_i$ and $\Xi_j$. $V_1$ determines the interaction of
\begin{align}
 \dot \matq \dot \matq (x)\rightarrow \dot \matq \dot \matq (y) &= -4 V_1(x,y)\ , \\
\dot \matq^+ \matq^+ (x)\rightarrow \dot \matq^+ \matq^+ (y) &= 2 V_1(x,y)\ , \\ \dot \matq \dot \matq (x)\rightarrow \dot \matq^+ \matq^+ (y) &= -2 \sqrt{2} i V_1(x,y)\ ,
\end{align}
while $V_2$ yields
\begin{equation}
\label{V2}
 \dot \matq \matq^+ (x)\rightarrow  \dot \matq \matq^+ (y) = V_2(x,y)\ .
\end{equation}
These are the only vertices that do not vanish for minimal coupling\cite{Bergamin:2004us}, $K = 1$. Both of them are conformally invariant, but while while $V_1$ vanishes at $x^0=y^0$, $V_2$ does not unless $K=1$, which thus yields a local four-point interaction for non-minimal coupling. The vertex function $V_1$ with minimal coupling vanishes for models with $w\propto\phi$, in particular for the CGHS model.\cite{Callan:1992rs} Since $V_1$ is the only vertex function of bosonic models with minimal couplings, the CGHS model exhibits scattering triviality to this order. This does not apply to the supersymmetric extension of the CGHS model\cite{Park:1993sd} since $V_2$ does not vanish here.\cite{Bergamin:2004us}

$V_3$ appears in two different interaction terms in \eqref{vertint} leading to the four vertices
\begin{align}
 \dot \matq \dot \matq (x)\rightarrow \partial \matq \dot \matq (y) &= 2i V_3(x,y)\ , & 
\dot \matq^+ \matq^+ (x)\rightarrow \partial \matq \dot \matq (y) &= -\sqrt{2} V_3(x,y)\ , \\ 
\dot \matq \dot \matq (x)\rightarrow \partial \matq^+ \matq^+ &= 2 \sqrt{2} V_3(x,y)\ , & 
\dot \matq^+ \matq^+ (x)\rightarrow \partial \matq^+ \matq^+ &= 2i V_3(x,y)\ .
\end{align}
The remaining two interactions with $(2i \Phi_0 - \sqrt{2} \Psi_0^{++)})$ as initial or final state are
\begin{align}
 \dot \matq \dot \matq (x)\rightarrow \matq^- \matq^+ (y) &=2iV_4(x,y)\ , & 
\dot \matq^+ \matq^+ (x)\rightarrow \matq^- \matq^+ (y) &= -\sqrt{2} V_4(x,y)\ , \\ 
\dot \matq \dot \matq (x)\rightarrow \dot \matq^- \matq^- (y)&= 2i V_5 (x,y)\ , &
\dot \matq^+ \matq^+ (x)\rightarrow \dot \matq^- \matq^- (y)&= -\sqrt{2} V_5 (x,y)\ .
\end{align}
Notice that these vertices are not conformally invariant\footnote{It might come as a surprise that a conformally invariant model generates vertices which are not invariant under this transformation. Nonetheless, it should be remembered that conformal invariance applies to the complete Lagrangian, here \eqref{vertint}, including the (asymptotic) matter states.\cite{Grumiller:2006ja,Grumiller:2006rc}} as they cannot be written exclusively in terms of the conformally invariant potential $w$. Finally, for non-minimal coupling besides \eqref{V2} there exist two additional vertices with mixed bosonic/fermionic initial and final states,
\begin{align}
\dot \matq \matq^+ (x) \rightarrow \dot \matq \matq^-(y) &= V_6 (x,y)\,, \\ 
\dot \matq \matq^+ (x) \rightarrow \partial \matq \matq^-(y) &= V_7 (x,y)\,.
\end{align}
$V_6$ is not conformally invariant but zero for $x^0=y^0$, while $V_7$ exactly behaves the other way around.

%% file: part3.tex
\section{Two-Dimensional Dilaton Gravity with Boundaries}\label{sect:boundary}
In all calculations of the previous sections boundary terms were assumed to vanish and asymptotically the values of the target space variables $\phi$ and $X^I$ were fixed, which removes eventual boundary degrees of freedom. However, in many applications this setup is not suitable and a careful treatment of boundary terms is necessary. Already in Ref.~\refcite{Kummer:1997si} it was found by Wolfgang Kummer and Stephen Lau that the action \eqref{eq:SFOG} should be complemented by the boundary term
\begin{equation}
\label{FOYGH}
 \Stext{boundary} = \int_{\partial \mathcal{M}} \left(X\omega + \frac12 X \extd \ln\left| \frac{e^+_{\|}}{e^-_{\|}}\right|\right)\,.
\end{equation}
in order to make the theory globally equivalent to the model \eqref{eq:SGDT} complemented with the standard York-Gibbons-Hawking boundary term \cite{York:1972sj,Gibbons:1977ue}. If Dirichlet boundary conditions are chosen for $X$, $e^+_{\|}$ and $e^-_{\|}$ the second term in Eq.~\eqref{FOYGH} is formulated exclusively in terms of fields fixed at the boundary. However, this term is essential to restore invariance under unrestricted Lorentz transformations.\footnote{In order to ensure a well-definedness of the semiclassical approximation and thus of thermodynamics of black hole spacetimes, further boundary counterterms, solely depending on the boundary values of the fields held fixed there, can be important. In the Euclidean approach these counterterms have been discussed in Refs.~\refcite{Grumiller:2007ju,Bergamin:2007sm}. 
}

In one of his last publications\cite{Bergamin:2005pg} Wolfgang Kummer resumed the discussion of boundary terms in FOG from a quite different point of view. This work was motivated by results from Refs.~\refcite{Carlip:2004mn,Carlip:2005xy} where it was argued that black hole entropy should emerge from Goldstone-like degrees of freedom that emerge from a symmetry breaking in the presence of a horizon.\footnote{This is thought to be a spontaneous symmetry breaking happening in the full dynamical theory. However, because of the inability to treat the fully dynamical picture, the investigation of Carlip's idea is done by implementing the symmetry breaking explicitly through boundary constraints.} In these works a stretched horizon was imposed as a boundary, implemented by suitable boundary constraints. It was then found that these constraints break parts of the symmetry, which allowed to deduce the correct entropy by means of the Cardy formula\cite{Cardy:1986ie}. Since the formalism of Refs.~\refcite{Carlip:2004mn,Carlip:2005xy} does not allow to impose sharp horizon constraints, it however remained open in which sense the stretched horizon really is a special choice of a boundary. In FOG the Eddington-Finkelstein type solutions are not singular at the horizon and thus the first order formulation provides the possibility to replace the stretched horizon in Refs.~\refcite{Carlip:2004mn,Carlip:2005xy} by a true horizon. This led to the idea to study FOG with boundaries, once chosen as a generic boundary and once chosen as a horizon, and to compare these two situations.

In Ref.~\refcite{Bergamin:2005pg} the boundary was considered at a fixed value of $x^1$, whereby $x^0$ represents Hamiltonian time. This choice allowed to implement the specific values of the fields fixed at the boundary as boundary constraints, which then were introduced in the Hamiltonian analysis. Since the dilaton is constant at the horizon, the first boundary constraint was chosen as $\hat B_1[\eta]=(p_1-p_1^b)\eta\vert_{\partial\mathcal{M}}$. A generic boundary is determined by the two additional constraints ($\eta$ is a smearing function)
\begin{align}
\hat B_2[\eta]&=(\bar{q}_2-E_0^-(x^0))\eta\vert_{\partial\mathcal{M}}\ , &
\hat B_3[\eta]&=(\bar{q}_3-E_0^+(x^0))\eta\vert_{\partial\mathcal{M}}\ ,\label{eq:hor0.33}
\end{align}
which turns all secondary constraints $G^I$ into second class constraints. On-shell \eqref{eq:hor0.33} with the choice $E_0^-(x^0) \equiv 0$ could be used to fix the boundary to be a horizon. However, off-shell this choice is problematic since the Killing norm expressed in terms of target space variables, Eq.~\eqref{eq:Killingnorm}, need not vanish and not surprisingly it was found that the Hamiltonian treatment of \eqref{eq:hor0.33} becomes singular at the horizon. Still, the first order formulation offers a different set of horizon constraints, namely
\begin{align}
B_2[\eta]&=\bar{q}_2\eta\vert_{\partial\mathcal{M}}\ , &
B_3[\eta]&=p_3\eta\vert_{\partial\mathcal{M}}\ ,\label{eq:hor0.3}
\end{align}
which removes all the problems encountered with \eqref{eq:hor0.33}. As an important difference to the generic boundary it is now found that two of the three secondary constraints, namely the Lorentz constraints and diffeomorphisms along the boundary, remain first class. This picture was confirmed by constructing the reduced phase space. Both situations have zero physical degrees of freedom in the bulk, but while a generic boundary exhibits one pair of boundary degrees of freedom (which could be related to mass and proper time as previously found by Kucha{\v r}\cite{Kuchar:1994zk}), no boundary degrees of freedom are left at the horizon. This suggests a quite different picture of black hole entropy\cite{Bergamin:2005pg,Bergamin:2006zy}: The physical degrees of freedom present on a generic boundary are converted into gauge degrees of freedom on a horizon and entropy arises because approaching the black hole horizon does not commute with constructing the physical phase space.

Already during the preparation of Ref.~\refcite{Bergamin:2005pg} it was realized by Wolfgang Kummer and his collaborators that it could be advantageous to choose the boundary at constant value of Hamiltonian time. Indeed, the extremely complex constraint algebras emerging from the calculations above destroyed any hopes to quantize the model along the lines of Sect.~\ref{sect:exactpathintegral}, not to mention the impossibility to couple matter fields. Nonetheless, if in the Hamiltonian picture the boundary is rather chosen as initial or final values, the canonical formalism is not affected at all. However, since for a spacelike boundary the boundary values are no longer fixed via boundary constraints, the necessary restrictions should be obtained from the ``lost constraints,''\footnote{Integrating out the $p_I$ reproduces only the equations of motion \eqref{eq:delta1}-\eqref{eq:delta3}. There is another set of these equations with spatial rather than time-derivatives, which should fix the remaining freedom in the choice of certain integration functions (depending on $x^1$). These are the ``lost constraints'', which play the role of Ward identities for the diffeomorphism and local Lorentz invariance. See also Ref.~\refcite{Grumiller:2001ea} and references therein.} which turned out to be difficult to tackle. Thus this line of investigations was given up in favor of the picture presented in Ref.~\refcite{Bergamin:2005pg}. Only recently, this unfinished work was continued\cite{Bergamin:2007aw} and it was shown that for the matterless theory the result expected from previous works\cite{Kuchar:1994zk,Bergamin:2005pg}, namely the existence of one boundary degree of freedom -- the mass -- was obtained. In particular, fixing $e_1^{\pm}$ at the boundary instead of $X^{\pm}$ implies for the arguments of the functional $\delta$ functions \eqref{eq:delta1}-\eqref{eq:delta3} not to evaluate the path integral completely, but rather leaving an integration over field boundary values unevaluated. Furthermore, additional contributions from the Gibbons-Hawking boundary term made the evaluation the of quantum equations of motion and the identification of the additional boundary degree of freedom possible. The formalism presented in Ref.~\refcite{Bergamin:2007aw} thus may provide a way to finally do a path integration over the remaining boundary degree of freedom -- a real ``sum over boundary conditions'' labeled by the mass of the spacetime. Though relevant questions regarding the path integral remained open this work, and will hopefully addressed in the future, it shows that Wolfgang Kummer's philosophy of the ``Vienna School of dilaton gravity'' remains a powerful formalism which also in the future will provide deeper insight into important questions of classical and quantum gravity.

%% file: appendix.tex
\label{sec:notation}
Most of the notation follows the one used in Refs.~
\refcite{Ertl:2000si,Ertl:2001sj}, which should be consulted for further explanations.

For indices of target-space coordinates and gauge fields the notation
\begin{gather}
 X^I = (X^i,X^\alpha) = (X^\phi,X^a,X^\alpha) = (\phi, X^a, \chi^\alpha)\ , \\
 A_I = (A_i,A_\alpha) = (A_\phi,A_a,A_\alpha) = (\omega, e_a, \psi_\alpha)\ ,
\end{gather}
i.e. capital Latin indices are generic, $i,j,k\ldots$ are bosonic, $a,b,c\ldots$ denote the anholonomic coordinates and Greek indices are fermionic. The summation convention is always $NW \rightarrow SE$, e.g.\ for a fermion $\chi$: $\chi^2 = \chi^\alpha \chi_\alpha$. Our conventions are
arranged in such a way that almost every bosonic expression is transformed
trivially to the graded case when using this summation convention and
replacing commuting indices by general ones. This is possible together with
exterior derivatives acting \emph{from the right}, only. Thus the graded
Leibniz rule is given by
\begin{equation}
  \label{eq:leibniz}
  \mbox{d}\left( AB\right) =A\mbox{d}B+\left( -1\right) ^{B}(\mbox{d}A) B\ .
\end{equation}

In terms of anholonomic indices the metric and the symplectic $2 \times 2$
tensor are defined as
\begin{align}
  \eta_{ab} &= \left( \begin{array}{cc} 1 & 0 \\ 0 & -1
  \end{array} \right)\ , &
  \epsilon_{ab} &= - \epsilon^{ab} = \left( \begin{array}{cc} 0 & 1 \\ -1 & 0
  \end{array} \right)\ , & \epsilon_{\alpha \beta} &= \epsilon^{\alpha \beta} = \left( \begin{array}{cc} 0 & 1 \\ -1 & 0
  \end{array} \right)\ .
\end{align}
The metric in terms of holonomic indices is obtained by $g_{mn} = e_n^b e_m^a
\eta_{ab}$ and for the determinant the standard expression $e = \det e_m^a =
\sqrt{- \det g_{mn}}$ is used. The volume form reads $\epsilon = \frac{1}{2}
\epsilon^{ab} e_b \wedge e_a$; by definition $\ast \epsilon = 1$.

The $\gamma$-matrices used are in a chiral representation:
\begin{align}
\label{eq:gammadef}
  {{\gamma^0}_\alpha}^\beta &= \left( \begin{array}{cc} 0 & 1 \\ 1 & 0
  \end{array} \right) & {{\gamma^1}_\alpha}^\beta &= \left( \begin{array}{cc} 0 & 1 \\ -1 & 0
  \end{array} \right) & {{\gthree}_\alpha}^\beta &= {(\gamma^1
    \gamma^0)_\alpha}^\beta = \left( \begin{array}{cc} 1 & 0 \\ 0 & -1
  \end{array} \right)
\end{align}

Covariant derivatives of anholonomic indices with respect to the geometric
variables $e_a = \extd x^m e_{am}$ and $\psi_\alpha = \extd x^m \psi_{\alpha m}$
include the two-dimensional spin-connection one form $\omega^{ab} = \omega
\epsilon^{ab}$. When acting on lower indices the explicit expressions read
($\frac{1}{2} \gthree$ is the generator of Lorentz transformations in spinor space):
\begin{align}
\label{eq:A8}
  (D e)_a &= \extd e_a + \omega {\epsilon_a}^b e_b & (D \psi)_\alpha &= \extd
  \psi_\alpha - \frac{1}{2} {{\omega \gthree}_\alpha}^\beta \psi_\beta
\end{align}

For Majorana spinors in chiral representation,
\begin{align}
\label{eq:Achi}
  \chi^\alpha &= ( \chi^+, \chi^-)\ , & \chi_\alpha &= \begin{pmatrix} \chi_+ \\
  \chi_- \end{pmatrix}\ ,
\end{align}
upper and lower chiral components are related by $\chi^+
= \chi_-$, $ \chi^- = - \chi_+$, $\chi^2 = \chi^\alpha \chi_\alpha = 2 \chi_- \chi_+$. Vectors conveniently are used in the spin tensor decomposition $v^{\alpha \beta} = \frac{i}{\sqrt{2}} v^c \gamma_c^{\alpha
  \beta}$. Due to the additional factor $i$ the spin tensor components are related to standard light cone components as
\begin{align}
\label{lightcone}
 v^{++} &= i v^{\oplus}\ , & v^{--} &= -i v^{\ominus}\ ,
\end{align}
in particular the spin tensor components of a real vector are imaginary.
This notation implies that $\eta_{++|--} = 1$, $\epsilon_{--|++} = - \epsilon_{++|--} = 1$ and for the $\gamma$ matrices one finds
\begin{align}
\label{eq:gammalc}
  {(\gamma^{++})_\alpha}^\beta &= \sqrt{2} i \left( \begin{array}{cc} 0 & 1 \\ 0 & 0
  \end{array} \right)\ , & {(\gamma^{--})_\alpha}^\beta &= - \sqrt{2} i \left( \begin{array}{cc} 0 & 0 \\ 1 & 0
  \end{array} \right)\ .
\end{align}

%% file: WolfgangMemorial.bbl
\begin{thebibliography}{103}
\providecommand{\natexlab}[1]{#1}
\providecommand{\url}[1]{\texttt{#1}}
\expandafter\ifx\csname urlstyle\endcsname\relax
  \providecommand{\doi}[1]{doi: #1}\else
  \providecommand{\doi}{doi: \begingroup \urlstyle{rm}\Url}\fi

\bibitem{Kummer:1991ss}
W.~Kummer and D.~J. Schwarz, {Classical and quantum theory of non-Einsteinian
  2d- gravity}, \emph{Czech. J. Phys.} {\bf 41}, \penalty0 13--22,  (1991).

\bibitem{Kummer:1991bt}
W.~Kummer and D.~J. Schwarz.
\newblock {NonEinsteinian gravity with torsion at d = 2}.
\newblock In \emph{Strings and Symmetries, 1991: SUNY Stony Brook, 20-25 May
  1991}, pp. 168--169. World Scientific,  (1992).

\bibitem{Kummer:1991bg}
W.~Kummer and D.~J. Schwarz, {General analytic solution of R**2 gravity with
  dynamical torsion in two-dimensions}, \emph{Phys. Rev.} {\bf D45}, \penalty0
  3628--3635,  (1992).

\bibitem{Kummer:1991rt}
W.~Kummer and D.~J. Schwarz, {Renormalization of R**2 gravity with dynamical
  torsion in d = 2}, \emph{Nucl. Phys.} {\bf B382}, \penalty0 171--186,
  (1992).

\bibitem{Katanaev:1986wk}
M.~O. Katanaev and I.~V. Volovich, String model with dynamical geometry and
  torsion, \emph{Phys. Lett.} {\bf B175}, \penalty0 413--416,  (1986).

\bibitem{Katanaev:1990qm}
M.~O. Katanaev and I.~V. Volovich, Two-dimensional gravity with dynamical
  torsion and strings, \emph{Ann. Phys.} {\bf 197}, \penalty0 1,  (1990).

\bibitem{Katanaev:1989mk}
M.~O. Katanaev, Complete integrability of two-dimensional gravity with
  dynamical torsion, \emph{J. Math. Phys.} {\bf 31}, \penalty0 882,  (1990).

\bibitem{Kummer:1974ze}
W.~Kummer, {Ghost Free Nonabelian Gauge Theory}, \emph{Acta Phys. Austriaca}.
  {\bf 41}, \penalty0 315--334,  (1975).

\bibitem{Grosse:1992eb}
H.~Grosse, W.~Kummer, P.~Presnajder, and D.~J. Schwarz, {Nonlinear gauge
  symmetry in R**2 gravity in two- dimensions}, \emph{Czech. J. Phys.} {\bf
  42}, \penalty0 1325--1329,  (1992).

\bibitem{Kummer:1992af}
W.~Kummer and D.~J. Schwarz, {Two-dimensional R**2 gravity with torsion},
  \emph{Class. Quant. Grav.} {\bf 10}, \penalty0 S235--S238,  (1993).

\bibitem{Grosse:1992vc}
H.~Grosse, W.~Kummer, P.~Presnajder, and D.~J. Schwarz, {Novel symmetry of
  nonEinsteinian gravity in two- dimensions}, \emph{J. Math. Phys.} {\bf 33},
  \penalty0 3892--3900,  (1992).

\bibitem{Kummer:1992ef}
W.~Kummer.
\newblock Deformed {ISO}(2,1) symmetry and non-{E}insteinian 2d-gravity with
  matter.
\newblock In eds. D.~Bruncko and J.~Urban, \emph{Hadron Structure '92},
  (1992).
\newblock Stara Lesna, Czechoslovakia.

\bibitem{Kawai:1993kp}
H.~Kawai and R.~Nakayama, {Quantum R**2 gravity in two-dimensions}, \emph{Phys.
  Lett.} {\bf B306}, \penalty0 224--232,  (1993).

\bibitem{Strobl:1993yn}
T.~Strobl, {Quantization and the issue of time for various two- dimensional
  models of gravity}, \emph{Int. J. Mod. Phys.} {\bf D3}, \penalty0 281--284,
  (1994).

\bibitem{Cangemi:1992bj}
D.~Cangemi and R.~Jackiw, {Gauge invariant formulations of lineal gravity},
  \emph{Phys. Rev. Lett.} {\bf 69}, \penalty0 233--236,  (1992).

\bibitem{Verlinde:1991rf}
H.~Verlinde.
\newblock Black holes and strings in two dimensions.
\newblock In \emph{Trieste Spring School on Strings and Quantum Gravity}, pp.
  178--207 (April, 1991).

\bibitem{Teitelboim:1983ux}
C.~Teitelboim, Gravitation and {H}amiltonian structure in two space-time
  dimensions, \emph{Phys. Lett.} {\bf B126}, \penalty0 41,  (1983).

\bibitem{jackiw:1984}
R.~Jackiw.
\newblock Liouville field theory: a two-dimensional model for gravity.
\newblock In ed. S.~Christensen, \emph{Quantum theory of gravity : essays in
  honor of the 60th birthday of Bryce S.DeWitt}, pp. 327--344, Bristol,
  (1984). Hilger.

\bibitem{Ikeda:1992qz}
N.~Ikeda and K.~I. Izawa, {Quantum gravity with dynamical torsion in two-
  dimensions}, \emph{Prog. Theor. Phys.} {\bf 89}, \penalty0 223--230,  (1993).

\bibitem{Ikeda:1993nk}
N.~Ikeda and K.~I. Izawa, {Gauge theory based on quadratic Lie algebras and 2-d
  gravity with dynamical torsion}, \emph{Prog. Theor. Phys.} {\bf 89},
  \penalty0 1077--1086,  (1993).

\bibitem{Grumiller:2002nm}
D.~Grumiller, W.~Kummer, and D.~V. Vassilevich, Dilaton gravity in two
  dimensions, \emph{Phys. Rept.} {\bf 369}, \penalty0 327,  (2002).

\bibitem{Kummer:1994ur}
W.~Kummer and P.~Widerin, Non{E}insteinian gravity in d=2: Symmetry and current
  algebra, \emph{Mod. Phys. Lett.} {\bf A9}, \penalty0 1407--1414,  (1994).

\bibitem{Katanaev:1994qf}
M.~O. Katanaev, Canonical quantization of the string with dynamical geometry
  and anomaly free nontrivial string in two- dimensions, \emph{Nucl. Phys.}
  {\bf B416}, \penalty0 563--605,  (1994).

\bibitem{Kummer:1993uk}
W.~Kummer.
\newblock {Exact classical and quantum integrability of R**2 + T**2 gravity in
  (1+1) dimensions}.
\newblock In eds. J.~Carr and M.~Perrottet, \emph{International Europhysics
  Conference On High-Energy Physics (HEP93)}. Editions Frontieres,  (1993).

\bibitem{Haider:1994cw}
F.~Haider and W.~Kummer, {Quantum functional integration of nonEinsteinian
  gravity in d = 2}, \emph{Int. J. Mod. Phys.} {\bf A9}, \penalty0 207--220,
  (1994).

\bibitem{Schaller:1994np}
P.~Schaller and T.~Strobl, {Canonical quantization of non-Einsteinian gravity
  and the problem of time}, \emph{Class. Quant. Grav.} {\bf 11}, \penalty0
  331--346,  (1994).

\bibitem{Kummer:1995fk}
W.~Kummer.
\newblock {Unified treatment of all 1+1 dimensional gravitation models}.
\newblock In eds. J.~Lemonne, C.~{Vander Velde}, and F.~Verbeure,
  \emph{International Europhysics Conference On High Energy Physics (HEP 95)}.
  World Scientific,  (1995).

\bibitem{Kummer:1995qv}
W.~Kummer and P.~Widerin, Conserved quasilocal quantities and general covariant
  theories in two-dimensions, \emph{Phys. Rev.} {\bf D52}, \penalty0
  6965--6975,  (1995).

\bibitem{Kummer:1995qh}
W.~Kummer.
\newblock {General treatment of all 2d covariant models}.
\newblock In ed. S.~Moskaliuk, \emph{12th Hutsulian Workshop On Methods Of
  Mathematical Physics}. Hadronic Press,  (1995).

\bibitem{Katanaev:1996bh}
M.~O. Katanaev, W.~Kummer, and H.~Liebl, {Geometric Interpretation and
  Classification of Global Solutions in Generalized Dilaton Gravity},
  \emph{Phys. Rev.} {\bf D53}, \penalty0 5609--5618,  (1996).

\bibitem{Katanaev:1997ni}
M.~O. Katanaev, W.~Kummer, and H.~Liebl, On the completeness of the black hole
  singularity in 2d dilaton theories, \emph{Nucl. Phys.} {\bf B486}, \penalty0
  353--370,  (1997).

\bibitem{Kummer:1997si}
W.~Kummer and S.~R. Lau, Boundary conditions and quasilocal energy in the
  canonical formulation of all 1 + 1 models of gravity, \emph{Annals Phys.}
  {\bf 258}, \penalty0 37--80,  (1997).

\bibitem{Strobl:1994yk}
T.~Strobl.
\newblock \emph{Poisson structure induced field theories and models of 1+1
  dimensional gravity}.
\newblock PhD thesis, {T}echnische {U}niversit{\"a}t {W}ien,  (1994).

\bibitem{Strobl:1999wv}
T.~Strobl.
\newblock Gravity in two spacetime dimensions.
\newblock Habilitation thesis,  (1999).

\bibitem{Ertl:2000si}
M.~Ertl, W.~Kummer, and T.~Strobl, General two-dimensional supergravity from
  {P}oisson superalgebras, \emph{JHEP}. {\bf 01}, \penalty0 042,  (2001).

\bibitem{Kummer:1997hy}
W.~Kummer, H.~Liebl, and D.~V. Vassilevich, Exact path integral quantization of
  generic 2-d dilaton gravity, \emph{Nucl. Phys.} {\bf B493}, \penalty0
  491--502,  (1997).

\bibitem{Balasin:2004gf}
H.~Balasin, C.~G. Boehmer, and D.~Grumiller, {The spherically symmetric
  standard model with gravity}, \emph{Gen. Rel. Grav.} {\bf 37}, \penalty0
  1435--1482,  (2005).

\bibitem{Klosch:1996fi}
T.~Kl{\"o}sch and T.~Strobl, Classical and quantum gravity in (1+1)-dimensions.
  {P}art 1: {A} unifying approach, \emph{Class. Quant. Grav.} {\bf 13},
  \penalty0 965--984,  (1996).

\bibitem{Klosch:1996qv}
T.~Kl{\"o}sch and T.~Strobl, Classical and quantum gravity in 1+1 dimensions,
  part {II}: {T}he universal coverings, \emph{Class. Quant. Grav.} {\bf 13},
  \penalty0 2395--2422,  (1996).

\bibitem{Ikeda:1994fh}
N.~Ikeda, {Two-dimensional gravity and nonlinear gauge theory}, \emph{Ann.
  Phys.} {\bf 235}, \penalty0 435--464,  (1994).

\bibitem{Schaller:1994uj}
P.~Schaller and T.~Strobl.
\newblock {Poisson sigma models: A generalization of 2-d gravity Yang- Mills
  systems}.
\newblock In \emph{Finite dimensional integrable systems}, pp. 181--190,
  (1994).
\newblock Dubna.

\bibitem{Schaller:1994es}
P.~Schaller and T.~Strobl, Poisson structure induced (topological) field
  theories, \emph{Mod. Phys. Lett.} {\bf A9}, \penalty0 3129--3136,  (1994).

\bibitem{Kummer:1997jj}
W.~Kummer, H.~Liebl, and D.~V. Vassilevich, {Non-perturbative path integral of
  2d dilaton gravity and two-loop effects from scalar matter}, \emph{Nucl.
  Phys.} {\bf B513}, \penalty0 723--734,  (1998).

\bibitem{Kummer:1998zs}
W.~Kummer, H.~Liebl, and D.~V. Vassilevich, Integrating geometry in general 2d
  dilaton gravity with matter, \emph{Nucl. Phys.} {\bf B544}, \penalty0
  403--431,  (1999).

\bibitem{Fischer:2001vz}
P.~Fischer, D.~Grumiller, W.~Kummer, and D.~V. Vassilevich, S-matrix for s-wave
  gravitational scattering, \emph{Phys. Lett.} {\bf B521}, \penalty0 357--363,
  (2001).
\newblock Erratum ibid. {\bf B532} (2002) 373.

\bibitem{Grumiller:2001rg}
D.~Grumiller, Virtual black hole phenomenology from 2d dilaton theories,
  \emph{Class. Quant. Grav.} {\bf 19}, \penalty0 997--1009,  (2002).

\bibitem{Bergamin:2004aw}
L.~Bergamin.
\newblock Quantum dilaton supergravity in 2d with non-minimally coupled matter.
\newblock In eds. P.~Fiziev and M.~Todorov, \emph{Gravity, Astrophysics, and
  Strings @ the Black Sea}, pp. 17--28, Sofia,  (2005). St.Kliment Ohridski
  University Press.

\bibitem{Grumiller:2006ja}
D.~Grumiller and R.~Meyer, {Quantum dilaton gravity in two dimensions with
  fermionic matter}, \emph{Class. Quant. Grav.} {\bf 23}, \penalty0 6435--6458,
   (2006).

\bibitem{Meyer:2005fz}
R.~Meyer.
\newblock {Constraints in two-dimensional dilaton gravity with fermions}.
\newblock To appear in the proceedings of International V.A. Fock School of
  Advances of Physics (IFSAP 2005), St. Petersburg, Russia, 21-27 Nov 2005.,
  (2005).

\bibitem{Meyer:2006vha}
R.~Meyer.
\newblock {Classical and quantum dilaton gravity in two dimensions with
  fermions}.
\newblock Master's thesis,  (2006).

\bibitem{Meyer:2006xm}
R.~Meyer.
\newblock {Quantizing two-dimensional dilaton gravity with fermions: The Vienna
  way}.
\newblock In eds. H.~Kleinert and R.~Jantzen, \emph{The eleventh {M}arcel
  {G}rossman meeting}. World Scientific,  (2008).

\bibitem{Bergamin:2002ju}
L.~Bergamin and W.~Kummer, Graded {P}oisson sigma models and dilaton-deformed
  2d supergravity algebra, \emph{JHEP}. {\bf 05}, \penalty0 074,  (2003).

\bibitem{Bergamin:2004us}
L.~Bergamin, D.~Grumiller, and W.~Kummer, Quantization of 2d dilaton
  supergravity with matter, \emph{JHEP}. {\bf 05}, \penalty0 060,  (2004).

\bibitem{Fradkin:1975cq}
E.~S. Fradkin and G.~A. Vilkovisky, Quantization of relativistic systems with
  constraints, \emph{Phys. Lett.} {\bf B55}, \penalty0 224,  (1975).

\bibitem{Fradkin:1978xi}
E.~S. Fradkin and T.~E. Fradkina, Quantization of relativistic systems with
  boson and fermion first and second class constraints, \emph{Phys. Lett.} {\bf
  B72}, \penalty0 343,  (1978).

\bibitem{Batalin:1977pb}
I.~A. Batalin and G.~A. Vilkovisky, Relativistic {S} matrix of dynamical
  systems with boson and fermion constraints, \emph{Phys. Lett.} {\bf B69},
  \penalty0 309--312,  (1977).

\bibitem{Grumiller:2001ea}
D.~Grumiller.
\newblock \emph{Quantum dilaton gravity in two dimensions with matter}.
\newblock PhD thesis, {T}echnische {U}niversit{\"a}t {W}ien,  (2001).

\bibitem{Grumiller:2002dm}
D.~Grumiller, W.~Kummer, and D.~V. Vassilevich, Virtual black holes in
  generalized dilaton theories and their special role in string gravity,
  \emph{Eur. Phys. J.} {\bf C30}, \penalty0 135--143,  (2003).

\bibitem{Grumiller:2000ah}
D.~Grumiller, W.~Kummer, and D.~V. Vassilevich, The virtual black hole in 2d
  quantum gravity, \emph{Nucl. Phys.} {\bf B580}, \penalty0 438--456,  (2000).

\bibitem{Grumiller:2004yq}
D.~Grumiller, {Virtual black holes and the S-matrix}, \emph{Int. J. Mod. Phys.}
  {\bf D13}, \penalty0 1973--2002,  (2004).

\bibitem{Grumiller:2006rc}
D.~Grumiller and R.~Meyer, {Ramifications of lineland}, \emph{Turk. J. Phys.}
  {\bf 30}, \penalty0 349--378,  (2006).

\bibitem{Howe:1979ia}
P.~S. Howe, Super {W}eyl transformations in two-dimensions, \emph{J. Phys.}
  {\bf A12}, \penalty0 393--402,  (1979).

\bibitem{Park:1993sd}
Y.-C. Park and A.~Strominger, Supersymmetry and positive energy in classical
  and quantum two-dimensional dilaton gravity, \emph{Phys. Rev.} {\bf D47},
  \penalty0 1569--1575,  (1993).

\bibitem{Grumiller:2003dh}
D.~Grumiller.
\newblock Three functions in dilaton gravity: The good, the bad and the muggy.
\newblock Lectures given at 14th International Hutsulian Workshop on
  Mathematical Theories and their Physical and Technical Applications (Timpani
  - Mathyphys 2002), Chernivtsi, Ukraine,  (2002).

\bibitem{Bilal:1993ss}
A.~Bilal, Positive energy theorem and supersymmetry in exactly solvable quantum
  corrected 2-d dilaton gravity, \emph{Phys. Rev.} {\bf D48}, \penalty0
  1665--1678,  (1993).

\bibitem{Nojiri:1993sx}
S.~Nojiri and I.~Oda, Dilatonic supergravity in two-dimensions and the
  disappearance of quantum black hole, \emph{Mod. Phys. Lett.} {\bf A8},
  \penalty0 53--62,  (1993).

\bibitem{Ertl:1998ib}
M.~F. Ertl, M.~O. Katanaev, and W.~Kummer, Generalized supergravity in two
  dimensions, \emph{Nucl. Phys.} {\bf B530}, \penalty0 457--486,  (1998).

\bibitem{Ertl:2001sj}
M.~Ertl.
\newblock \emph{Supergravity in two spacetime dimensions}.
\newblock PhD thesis, {T}echnische {U}niversit{\"a}t {W}ien,  (2001).

\bibitem{Rivelles:1994xs}
V.~O. Rivelles, Topological two-dimensional dilaton supergravity, \emph{Phys.
  Lett.} {\bf B321}, \penalty0 189--192,  (1994).

\bibitem{Cangemi:1993mj}
D.~Cangemi and M.~Leblanc, {Two-dimensional gauge theoretic supergravities},
  \emph{Nucl. Phys.} {\bf B420}, \penalty0 363--378,  (1994).

\bibitem{Ikeda:1994dr}
N.~Ikeda, Gauge theory based on nonlinear {L}ie superalgebras and structure of
  2-d dilaton supergravity, \emph{Int. J. Mod. Phys.} {\bf A9}, \penalty0
  1137--1152,  (1994).

\bibitem{Izquierdo:1998hg}
J.~M. Izquierdo, Free differential algebras and generic 2d dilatonic
  (super)gravities, \emph{Phys. Rev.} {\bf D59}, \penalty0 084017,  (1999).

\bibitem{Strobl:1999zz}
T.~Strobl, Target-superspace in 2d dilatonic supergravity, \emph{Phys. Lett.}
  {\bf B460}, \penalty0 87--93,  (1999).

\bibitem{Bergamin:2004sk}
L.~Bergamin, Generalized complex geometry and the {P}oisson sigma model,
  \emph{Mod. Phys. Lett.} {\bf A20}, \penalty0 985--996,  (2005).

\bibitem{Calvo:2005ww}
I.~Calvo, {Supersymmetric WZ-Poisson sigma model and twisted generalized
  complex geometry}, \emph{Lett. Math. Phys.} {\bf 77}, \penalty0 53--62,
  (2006).

\bibitem{Strobl:2003kb}
T.~Strobl, Gravity from lie algebroid morphisms, \emph{Commun. Math. Phys.}
  {\bf 246}, \penalty0 475--502,  (2004).

\bibitem{Adak:2007xc}
M.~Adak and D.~Grumiller, {Poisson-sigma model for 2D gravity with
  non-metricity}, \emph{Class. Quant. Grav.} {\bf 24}, \penalty0 F65,  (2007).

\bibitem{Freedman:1976xh}
D.~Z. Freedman, P.~{van Nieuwenhuizen}, and S.~Ferrara, Progress toward a
  theory of supergravity, \emph{Phys. Rev.} {\bf D13}, \penalty0 3214--3218,
  (1976).

\bibitem{Freedman:1976py}
D.~Z. Freedman and P.~{van Nieuwenhuizen}, Properties of supergravity theory,
  \emph{Phys. Rev.} {\bf D14}, \penalty0 912,  (1976).

\bibitem{Deser:1976eh}
S.~Deser and B.~Zumino, Consistent supergravity, \emph{Phys. Lett.} {\bf B62},
  \penalty0 335,  (1976).

\bibitem{Grimm:1978kp}
R.~Grimm, J.~Wess, and B.~Zumino, Consistency checks on the superspace
  formulation of supergravity, \emph{Phys. Lett.} {\bf B73}, \penalty0 415,
  (1978).

\bibitem{Bergamin:2003am}
L.~Bergamin and W.~Kummer, The complete solution of 2d superfield supergravity
  from graded {P}oisson-sigma models and the super pointparticle, \emph{Phys.
  Rev.} {\bf D68}, \penalty0 104005,  (2003).

\bibitem{Bergamin:2003mh}
L.~Bergamin, D.~Grumiller, and W.~Kummer, Supersymmetric black holes in 2-d
  dilaton supergravity: baldness and extremality, \emph{J. Phys.} {\bf A37},
  \penalty0 3881--3901,  (2004).

\bibitem{Gibbons:1982fy}
G.~W. Gibbons and C.~M. Hull, A {B}ogomolny bound for general relativity and
  solitons in {N}=2 supergravity, \emph{Phys. Lett.} {\bf B109}, \penalty0 190,
   (1982).

\bibitem{Tod:1983pm}
K.~P. Tod, All metrics admitting supercovariantly constant spinors, \emph{Phys.
  Lett.} {\bf B121}, \penalty0 241--244,  (1983).

\bibitem{Nelson:1993vm}
W.~M. Nelson and Y.~Park, N=2 supersymmetry in two-dimensional dilaton gravity,
  \emph{Phys. Rev.} {\bf D48}, \penalty0 4708--4712,  (1993).

\bibitem{Bergamin:2004sr}
L.~Bergamin and W.~Kummer, Two-dimensional {N}=(2,2) dilaton supergravity from
  graded {P}oisson-sigma models {I}: Complete actions and their symmetries.,
  \emph{Eur. Phys. J.} {\bf C39}, \penalty0 S41--S52,  (2005).

\bibitem{Bergamin:2004na}
L.~Bergamin and W.~Kummer, Two-dimensional {N} = (2,2) dilaton supergravity
  from graded {P}oisson-sigma models. {II}: Analytic solution and {BPS} states,
  \emph{Eur. Phys. J.} {\bf C39}, \penalty0 S53--S63,  (2005).

\bibitem{Dirac:1996}
P.~A.~M. Dirac, \emph{{Lectures on Quantum Mechanics}}. (Belfer Graduate School
  of Science, Yeshiva University, New York, 1996).

\bibitem{Rocek:1986iz}
M.~Rocek, P.~{van Nieuwenhuizen}, and S.~C. Zhang, Superspace path integral
  measure of the n=1 spinning string, \emph{Ann. Phys.} {\bf 172}, \penalty0
  348,  (1986).

\bibitem{Lindstrom:1988fr}
U.~Lindstrom, N.~K. Nielsen, M.~Rocek, and P.~{van Nieuwenhuizen}, The
  supersymmetric regularized path integral measure in x space, \emph{Phys.
  Rev.} {\bf D37}, \penalty0 3588,  (1988).

\bibitem{Callan:1992rs}
J.~C.~G. Callan, S.~B. Giddings, J.~A. Harvey, and A.~Strominger, Evanescent
  black holes, \emph{Phys. Rev.} {\bf D45}, \penalty0 1005--1009,  (1992).

\bibitem{York:1972sj}
J.~W.~J. York, {Role of conformal three geometry in the dynamics of
  gravitation}, \emph{Phys. Rev. Lett.} {\bf 28}, \penalty0 1082--1085,
  (1972).

\bibitem{Gibbons:1977ue}
G.~W. Gibbons and S.~W. Hawking, Action integrals and partition functions in
  quantum gravity, \emph{Phys. Rev.} {\bf D15}, \penalty0 2752--2756,  (1977).

\bibitem{Grumiller:2007ju}
D.~Grumiller and R.~McNees, Thermodynamics of black holes in two (and higher)
  dimensions, \emph{JHEP}. {\bf 04}, \penalty0 074,  (2007).

\bibitem{Bergamin:2007sm}
L.~Bergamin, D.~Grumiller, R.~McNees, and R.~Meyer, {Black Hole Thermodynamics
  and Hamilton-Jacobi Counterterm}, \emph{J. Phys.} {\bf A41}, \penalty0
  164068,  (2008).

\bibitem{Bergamin:2005pg}
L.~Bergamin, D.~Grumiller, W.~Kummer, and D.~V. Vassilevich, Physics-to-gauge
  conversion at black hole horizons, \emph{Class. Quant. Grav.} {\bf 23},
  \penalty0 3075--3101,  (2006).

\bibitem{Carlip:2004mn}
S.~Carlip, {Horizon constraints and black hole entropy}, \emph{Class. Quant.
  Grav.} {\bf 22}, \penalty0 1303--1312,  (2005).

\bibitem{Carlip:2005xy}
S.~Carlip, {Horizon constraints and black hole entropy}.  (2005).

\bibitem{Cardy:1986ie}
J.~L. Cardy, {Operator Content of Two-Dimensional Conformally Invariant
  Theories}, \emph{Nucl. Phys.} {\bf B270}, \penalty0 186--204,  (1986).

\bibitem{Kuchar:1994zk}
K.~V. Kucha{\v{r}}, Geometrodynamics of {S}chwarzschild black holes,
  \emph{Phys. Rev.} {\bf D50}, \penalty0 3961--3981,  (1994).

\bibitem{Bergamin:2006zy}
L.~Bergamin and D.~Grumiller, {Killing horizons kill horizon degrees},
  \emph{Int. J. Mod. Phys.} {\bf D15}, \penalty0 2279--2284,  (2006).

\bibitem{Bergamin:2007aw}
L.~Bergamin and R.~Meyer.
\newblock Two-dimensional quantum gravity with boundary.
\newblock In eds. P.~Fiziev and M.~Todorov, \emph{Gravity, Astrophysics, and
  Strings @ the Black Sea}. St.Kliment Ohridski University Press,  (2008).

\end{thebibliography}
